\documentclass[a4paper,fleqn]{cas-sc}
\usepackage{diagbox}
\usepackage[authoryear,longnamesfirst]{natbib}
\usepackage{wrapfig}
\usepackage[ruled,vlined]{algorithm2e}
\usepackage{algpseudocode} 

\graphicspath{ {Figures/} }
\def\tsc#1{\csdef{#1}{\textsc{\lowercase{#1}}\xspace}}
\tsc{WGM}
\tsc{QE}
\tsc{EP}
\tsc{PMS}
\tsc{BEC}
\tsc{DE}

\begin{document}
\let\WriteBookmarks\relax
\def\floatpagepagefraction{1}
\def\textpagefraction{.001}
\shorttitle{HAV-DF: Deepfake}

\title [mode = title]{Hindi audio-video-Deepfake (HAV-DF): A Hindi language-based Audio-video Deepfake Dataset}                      
\author[1]{Sukhandeep Kaur}\corref{cor2}
\ead{sukhandeep.kaur@bmu.edu.in }
\author[1]{Mubashir Buhari}
\author[1]{Naman Khandelwal}
\author[1]{Priyansh Tyagi}
\author[1,2]{Kiran Sharma}\corref{cor1}
\ead{kiran.sharma@bmu.edu.in}

\address{School of Engineering \& Technology, BML Munjal University, Gurugram, Haryana-122413, India }
\address{Center for Advanced Data and Computational Science, BML Munjal University, Gurugram, Haryana-122413, India }
\cortext[cor2]{Principal author}
\cortext[cor1]{Corresponding author}


\begin{abstract}
Deepfakes, while technologically impressive, present a dual-edged sword. On the one hand, they have potential for innovation and creativity. However, their misuse poses significant risks to privacy, trust, and security, making it essential to develop detection tools and ethical frameworks to mitigate harm. With a vast Hindi-speaking population, India is particularly vulnerable to deepfake-driven misinformation campaigns. Fake videos or speeches in Hindi can have an enormous impact on rural and semi-urban communities, where digital literacy tends to be lower and people are more inclined to trust video content. The development of effective frameworks and detection tools to combat deepfake misuse requires high-quality, diverse, and extensive datasets. The existing popular datasets like FF-DF (FaceForensics++), and DFDC (DeepFake Detection Challenge) are based on English language, and there is a lack of regional language datasets. Hence, the aim of this paper is to create a first novel Hindi deep fake dataset, named ``Hindi audio-video-Deepfake'' (HAV-DF). The dataset has been generated using the faceswap, lipsyn and voice cloning methods. This multistep process allows us to create a rich, varied dataset that captures the nuances of Hindi speech and facial expressions, providing a robust foundation for training and evaluating deepfake detection models in a Hindi language context. It is unique of its kind as all of the previous datasets contain either deepfake videos or synthesized audio. This type of deepfake dataset can be used for training a detector for both deepfake video and  audio datasets. Notably, the newly introduced HAV-DF dataset demonstrates lower detection accuracy's across existing detection methods like Headpose, Xception-c40 Mesonet, etc. Compared to other well-known datasets FF-DF (FaceForensics++), and DFDC (DeepFake Detection Challenge). This trend suggests that the HAV-DF dataset presents deeper challenges to detect, possibly due to its focus on Hindi language content and diverse manipulation techniques. Furthermore, the HAV-DF dataset represents a significant contribution to the field, addressing the previous lack of Hindi-specific deepfake datasets and enabling more effective development of multilingual deepfake identification systems.
\end{abstract}

%
%
%

\begin{keywords}
Deepfake \sep Face swap  \sep Lipsync \sep Hindi dataset \sep Voice cloning

\end{keywords}

\maketitle

\section{Introduction}

Recent advances in AI, particularly in Generative AI and deep learning, have enabled the creation and manipulation of large volumes of highly realistic fake data across various formats, including audio, text, and video. The quality of these synthetic data is often indistinguishable from real data, making it challenging for human perception to differentiate between the two. Although these developments initially focused on textual data, the proliferation of social networks has facilitated the generation of fake content in other modalities, such as audio, video, and images. These visual and auditory formats spread misinformation more rapidly than text, as individuals tend to respond more quickly and strongly to visual content. Despite its potential for harm, synthetic data also offers significant benefits in fields such as education, entertainment, animation, and accessibility for visually impaired individuals~\citep{liz2024generation}.

A notable example of the negative implications of deepfakes occurred in 2017, when face-swapping techniques were used to create explicit videos featuring celebrities, which were then published online. Another significant case involved a fake video of former President Barack Obama, generated using software primarily developed by Reddit users \citep{mirsky2021creation}. Highlighting the seriousness of audio deepfake attacks, a recent cybercrime involving audio deepfakes led to a financial loss of \$35 million for a UAE-based company\citep{rabhi2024audio}.

The increasing interest in deepfakes research has led to the emergence of numerous workshops, conferences, and dedicated international projects, such as MediFor, funded by the Defense Advanced Research Projects Agency (DARPA). Furthermore, competitions such as the Media Forensics Challenge (MFC2018) and the Deepfake Detection Challenge (DFDC), initiated by the National Institute of Standards and Technology (NIST) and Facebook, have further driven advancements in the field \citep{tolosana2020deepfakes}. To develop an effective deepfake detection system, it is crucial to create large-scale high-quality data sets comprising authentic and manipulated media, allowing the distinction between real and synthetic content. Deepfakes are typically produced either by generating entirely synthetic data or by altering existing media. Training detection models on a diverse range of forgeries enhances their ability to generalize and perform well on previously unseen examples. This process also helps uncover specific features that characterize deepfakes, which are essential to designing robust detection systems \citep{zhou2017two,rossler2019faceforensics++}. Furthermore, creating benchmark detection models involves generating deepfakes with varying levels of complexity, ranging from basic modifications to highly advanced forgeries, to allow thorough evaluation and refinement of detection techniques \citep{dolhansky2020deepfake}.

Researchers can refine the deepfakes generated to assess the detection system's performance under different conditions, including compressed videos, low-resolution images, and noisy environments. With the continuous advancements in deepfake generation technologies, it is crucial to evaluate these systems against the most recent and sophisticated techniques to ensure the development of effective and up-to-date detection methods \citep{chesney2019deepfakes}. For audio-visual deepfakes, maintaining synchronization between audio and visual elements poses a major challenge, often leading to inconsistencies in lip-sync, facial expressions, or vocal patterns. Detecting such discrepancies during the creation process, such as misalignments between speech and facial movements, can improve the accuracy of future deepfake detection methods \citep{korshunov2018deepfakes}. The creation of deepfakes is essential for thoroughly evaluating the robustness of detection models. By producing forgeries with different degrees of realism, researchers can benchmark detection systems across a broad spectrum of manipulations \citep{dolhansky2020deepfake}. The most convincing deepfakes are often generated using diverse training datasets, efficient processes, and high-quality outputs that closely replicate real-world data.


\subsection{Deepfakes Types and Overview}

\begin{figure}[!ht]
    \centering
\includegraphics[width=\linewidth]{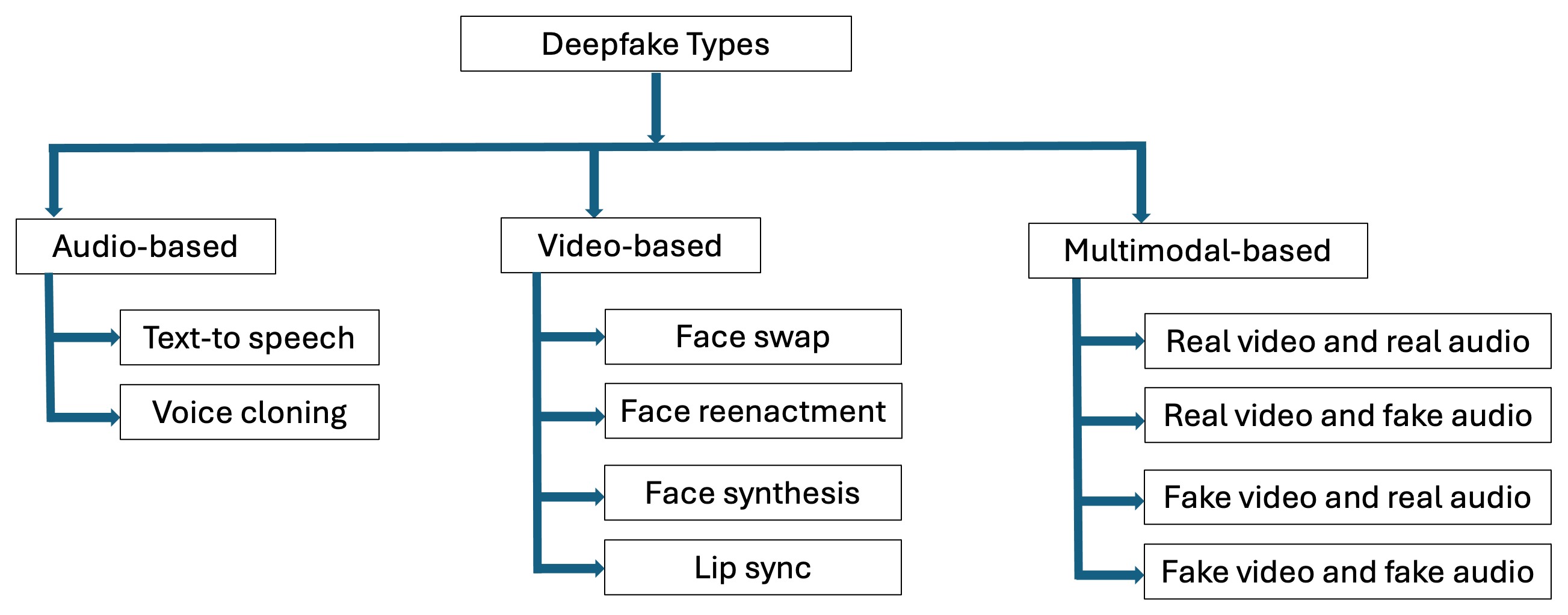} 
\caption{Types of Deepfakes.}
\label{fig:deepfakes}
\end{figure}

\cite{masood2023deepfakes} conducted an in-depth survey on deepfake generation and detection, categorizing deepfakes into two main types: audio and visual. For visual deepfakes, they identified five methodologies: Face Swap, Puppet-Mastery, Lip Syncing, Entire Face Synthesis, and Facial Attribute Manipulation. For audio deepfakes, they highlighted two primary techniques: Text-to-Speech Synthesis and Voice Conversion. Additionally, a multimodal category was noted, combining audio and visual deepfakes. Figure~\ref{fig:deepfakes} illustrates the various types of deepfakes generated. Mathematical representations use $s$ and $t$ to denote source and target identities, with $x_s$ and $x_t$ representing source and target images, respectively, and $x_g$ representing the generated deepfake created using source and target images. For generating HAV-DF dataset, we have considered the following deepfakes generation methods.

 \begin{enumerate}
     
  

\item \textbf{Face reenactment:} Face reenactment is a technique used to manipulate and transfer facial expressions, head movements, and gestures from one individual (source) to another (target) in a video or image. The goal is to make it appear as though the target person is performing the actions or expressions of the source while maintaining the target's identity. Puppet Master or Face Reenactment methods involve manipulating a person's facial features, such as head movements, eye motions, and gestures. These methods find extensive use in the animation and entertainment industries, particularly for tasks such as dubbing or modifying an actor's expressions in movie scenes. The mathematical process is outlined as follows:

    \begin{enumerate}
    
        \item \textit{Feature extraction}: Extract expression features from the source face \( x_s \) and identity features from the target face \( x_t \):
        \begin{equation}
   Z_{exp} = E_{exp}(x_s)
   \label{eq:exp}
   \end{equation}
   \begin{equation}
   Z_{id} = E_{id}(x_t)
   \label{eq:id}
   \end{equation}
   where \( Z_{exp} \) represents the source expressions and \( Z_{id} \) represents the target identity.

    \item \textit{Feature fusion}:
   - Combine the source’s expression features with the target’s identity features:
   \begin{equation}
   Z_{fusion} = f(Z_{exp}, Z_{id})
   \label{eq:fusion}
   \end{equation}
   Here, \( Z_{fusion} \) is the combined representation of the target's identity and source's expressions.

    \item \textit{Reenacted face generation}:
   - Generate the reenacted face \( x_g \) using a decoder \( D \):
   \begin{equation}
   x_g = D(Z_{fusion})
   \label{eq:generation}
   \end{equation}
    
\end{enumerate}

    \item \textbf{Face swap or replacement:} In Face Swap, the face from the source image $x_s$ is swapped with the face from the target video 
$x_t$. This process entails extracting the face from the source image and replacing key facial features, including the eyes, mouth, and nose of the target face, with those from the source. Various deepfake generation methods based on Face Swap have been examined in the literature, including Faceswap, FaceSwapGAN, DeepFaceLab, Fast Face Swap, FSNet, RSGAN, and FaceShifter. These approaches utilize architectures such as encoder-decoder networks, Generative Adversarial Networks (GANs), and Convolutional Neural Networks (CNNs) to create deepfakes. As described in Equation~\ref{eq:FS}, the contents of the target image $x_t$ are replaced with those of the source image $x_s$. The encoder $E$ maps the input image to a latent space vector. The process can be represented mathematically as follows:
\begin{enumerate}

    \item \textit{Encoding source and target faces:}
        \begin{equation}
        \centering
        Z_{s} = E(x_s), \quad Z_{t} = E(x_t)
        \label{eq:FS}
        \end{equation}
        where \( Z_{s} \) and \( Z_{t} \) are the latent features of the source \( x_s \) and target \( x_t \), extracted by the encoder \( E \).

        \item \textit{Combining features}:
        \begin{equation}
        \centering
        Z_{swap} = f(Z_{id,s}, Z_{attr,t})
        \end{equation}
        where \( Z_{id,s} \) represents the identity features of the source, \( Z_{attr,t} \) represents the attributes (e.g., pose, lighting) of the target, and \( f \) is the fusion function.
        
        \item \textit{Decoding to generate swapped face}:
        \begin{equation}
        \centering
        x_{swap} = D(Z_{swap})
        \end{equation}
        where \( D \) is the decoder that synthesizes the swapped face from the combined latent representation \( Z_{swap} \).
        
        \item \textit{Final composition}:
        \begin{equation}
        \centering
        x_{output} = B(x_{swap}, x_t)
        \end{equation}
        where \( B \) is the blending function that integrates the swapped face \( x_{swap} \) into the target image \( x_t \), preserving its background and context.
        
\end{enumerate}

    \item \textbf{Face Synthesis and Editing:} 
Face Synthesis and Editing involve generating new faces or modifying attributes of existing faces, such as age, gender, or expression. Face synthesis creates entirely new, realistic faces using models like GANs (e.g., StyleGAN). Face synthesis and editing involve generating new faces or modifying specific attributes of existing faces, such as age, expression, or hairstyle. These processes can be described mathematically as follows:
\begin{enumerate}
    \item \textit{Face synthesis}:
    \begin{equation}
    \centering
    x_g = G(z)
    \end{equation}
    where \( z \) is a latent vector sampled from a distribution \( P(z) \) (e.g., Gaussian), and \( G \) is the generator that outputs the synthesized face \( x_g \).
    
   \item \textit{Face editing}:
    \begin{equation}
    \centering
    x_{edit} = G(z + \Delta z_{attr})
    \end{equation}
    where \( z \) is the latent representation of the original image, \( \Delta z_{attr} \) represents the change in the latent space corresponding to a specific attribute (e.g., aging, expression), and \( x_{edit} \) is the edited face output.
    
    \item \textit{Attribute manipulation in latent space}:
    \begin{equation}
    \centering
    z_{edit} = z + \Delta z_{attr}
    \end{equation}
    This modifies the original latent representation \( z \) by adding a vector \( \Delta z_{attr} \), which corresponds to the desired change in attributes.
    \end{enumerate}

    
 
   

\item \textbf{Lip-synced:} Lip sync is the process of matching a speaker's lip movements in a video or animation to an audio track, such as spoken words or singing. The aim is to create a smooth and natural alignment between the visual lip movements and the audio, making it appear as though the person or character is authentically speaking the provided audio. The lip-synced process creates realistic lip movements in a video to match an audio input. The steps and mathematical equations are as follows:

    \begin{enumerate}
    
    \item \textit{Audio encoding}:
    The audio \( a \) is analyzed to extract speech features:
       \begin{equation}
         Z_{audio} = E_{audio}(a)
        \end{equation}
    where $Z_{audio}$ is latent representation of the audio and $E_{audio}$ is audio encoder.
    
    \item \textit{Video encoding}:
    The video frame \( x_t \) is analyzed to extract lip features:
       \begin{equation}
        Z_{lip} = E_{lip}(x_t)
     \end{equation}
    where $Z_{lip}$ is the latent representation of the visual features and $E_{lip}$ is visual encoder.
    
    \item \textit{Feature fusion}:
    The audio and lip features are combined to align lip movements with speech:
       \begin{equation}
        Z_{sync} = f(Z_{audio}, Z_{lip})
    \end{equation}
        where $Z_{sync}$ is the latent representation combining audio and visual features and $f$ is fusion function.
        
    \item \textit{Lip-synced frame generation}:
    A new video frame \( x_{sync} \) is created with synchronized lips:
      \begin{equation}
         x_{sync} = G(Z_{sync})
   \end{equation}

    where $x_{sync}$ is generated frame with lip-synced movements and $G$ is generator model.
    \end{enumerate}

The final equation is:
   \begin{equation}
    x_{sync} = G(f(E_{audio}(a), E_{lip}(x_t)))
\end{equation}

This process aligns audio and video features to generate a lip-synced video.


\item \textbf{Voice cloning:} Voice cloning is the process of replicating a person's voice using AI-based technologies. It involves capturing the unique characteristics of an individual’s voice—such as tone, pitch, accent, and speaking style—to synthesize new speech that mimics the original speaker. This technology is widely used in personalized virtual assistants, content creation, and accessibility tools. A sample of the target speaker’s voice is recorded. This sample can range from a few seconds (for zero-shot cloning) to several minutes for higher accuracy. This process can be described mathematically as follows:

\begin{enumerate}

    \item \textit{Feature extraction}: An encoder \( E_{voice} \) processes the input speech \( s \) to extract a speaker embedding \( Z_{voice} \), which captures the speaker's unique vocal characteristics:
    \begin{equation}
        Z_{voice} = E_{voice}(s)
    \end{equation}
    
    \item \textit{Text-to-speech synthesis}: A text input \( t \) is converted into a latent representation \( Z_{text} \) using a text encoder \( E_{text} \):
    \begin{equation}
        Z_{text} = E_{text}(t)
    \end{equation}
    
    \item \textit{Speech generation}: The voice embedding \( Z_{voice} \) and text representation \( Z_{text} \) are combined to generate the final audio output \( y \) using a vocoder \( V \):
    \begin{equation}
        y = V(f(Z_{voice}, Z_{text}))
    \end{equation}
    Here, \( f \) is a fusion function that aligns the speaker's vocal characteristics with the desired speech content.

    \end{enumerate}
 
\end{enumerate}

\subsection{Challenges in Deepfake Generation}
The major challenges in creating deepfakes are:
\begin{enumerate}
    \item The trade-off between data and quality: To get the better results we need huge amount of training data which can be feasible for some particular target or group of individuals like celebrities etc. but not feasible for individuals.
    \item Speed and quality: The quality of the deepfake is highly effected by the speed and it depends more on the channel of deepfake generation such as online (interactive) or offline (stored data). 
\item Availability and quality: The availability of more code and data will help to the researchers to generate high quality deepfakes\citep{mirsky2021creation}.
\end{enumerate}

\subsection{Research Objectives}
\begin{enumerate}
   
\item Create a Hindi-language based audio-video deepfake dataset named Hindi audio-video-Deepfake (HAV-DF).
\item Validate the robustness of the dataset using the pre-trained models.
\item Compare the performance of the HAV-DF dataset with the existing datasets on audio-video, only audio, and video only.

\end{enumerate}
\subsection{Deepfakes for Low Resource Languages}

Most research on audio-visual deepfakes has focused on high-resource languages such as English and Chinese, while low-resource languages like Indic, Arabic, and many Asian and European languages lack adequate attention. Low-resource languages are characterized by limited digital resources, including datasets, research, and technological support, which restricts advancements in deepfake detection methods for these languages. A significant challenge in generating deepfakes for low-resource languages is the lack of training data. Generating deepfakes necessitates extensive datasets of video and audio recordings to train neural networks on language-specific features. Low-resource languages, such as Indic languages, pose unique challenges due to their intricate grammatical structures, distinct phonetic characteristics, and culturally embedded visual expressions like gestures. The absence of language-specific or culturally tailored models significantly hampers the ability to produce accurate deepfakes.

A common approach to tackle this challenge is transfer learning, which involves fine-tuning a pre-trained model trained on high-resource languages for use with low-resource languages. This method leverages the generic features learned from extensive datasets and adapts them to the limited data available for low-resource languages \citep{jia2018transfer}. Voice cloning provides an effective solution for audio data in low-resource languages. By synthesizing speech data, it facilitates deepfake generation even with limited resources. Techniques such as few-shot learning enable models to accurately replicate a speaker's voice in a low-resource language using only a few samples \citep{arik2018neu}.

Multispeaker text-to-speech (TTS) models can be tailored for low-resource languages using cross-lingual training or language adaptation methods. These models are capable of synthesizing speech that accurately replicates real individuals' voices, even with limited data availability. However, implementing such solutions requires careful consideration of linguistic and cultural nuances to ensure accurate and ethical use of deepfake technology. For low-resource languages, it is crucial to involve the respective language communities in deepfake research to ensure ethical use of the technology. Addressing the unique challenges faced by low-resource languages not only enhances deep-fake detection capabilities, but also promotes equitable technological development.

\section{Literature Review}

\cite{mubarak2023survey} have done in depth survey on detection and impact of deepfakes with respect to visual, audio and text. They have examined the impacts of deepfakes in social, technical, political and economic domains. All deepfakes detection methods for audio, visual and textual have been analyzed separately for deep learning as well as handcrafted fetaure extraction methods. \cite{tolosana2020deepfakes} have reviewed the techniques for face manipulation to generate deepfakes. Liz-López et al have reviewed the generation and manipulation techniques for deepfakes in terms of forensic from 2018 to 2923 including the datasets. The in depth analysis of all the benchmarks datasets available for audio, visual, audio-visual has been given by~\cite{liz2024generation}. The features most of the researchers have used in the past for lip sync based deep fake generation are Mouth landmarks, MFCC audio features (28-D), VGG-M network, MFCC audio features, Mel-spectrogram representation etc. In Facial attribute manipulation, the particular facial attributes are manipulated such adding/removing eyeglasses, skin manipulation such as smoothening, removing scars, etc. \citep{masood2023deepfakes}. 

In audio manipulations, the deepfakes are generated by cloning a person's voice with some other voice which he never said before. Although synthesis voice is very helpful for development of entertainment industry, chatbots, AI assistant, assisting specially abele people. So the  two popular techniques available for audio synthesis are text to speech and voice cloning. In text to speech, it generates the synthetic voice similar to speaker's natural sound from a given input text while voice cloning works on manipulation of waveforms of the source speaker to convert it into target speaker waveforms~\citep{masood2023deepfakes}. 

Further the author had also mentioned the limitations of existing datasets and reviewed the lipsync based deep fake generation like LIpGAN, Wav2Lip, based on GAN models. The author had highlighted the many deepfake generation techniques based on face re-enactment such as Face2Face, ReenactGAN, GANimation, GANnotation, X2face, FaR-GAN etc. He further reviewed face synthesis deepfake generation technique CoGAN, ProGAN, StyleGAN, TP-GAN, SAGAN, BigGAN, StackGAN. Further, some researchers have done facial attribute manipulation base deepfake generation such as techniques used are IcGAN, StarGAN, AttGAN, STGAN, PA-GAN, etc. \citep{masood2023deepfakes}.
\cite{kwon2021kodf} have designed KODF Korean deepfake dataset using FaceSwap, DeepFaceLab, FSGAN,  FOMM i.e First Order Motion Model (FOMM) is a video-driven face-reenactment model, Wav2Lip . \cite{korshunov2018deepfakes} have evaluated the various deepfake videos generated using VGG and Facenet neural networks.
\subsection{Deepfake Generation of Videos}
Most of the available deepfake datasets have been generated using tools like StyleGAN, GANimation, FaceSwapping, SimSwap etc. For generation the foundation models used are autoencoder and GANS. Some of the best notable examples are: variational autoencoders, VAE-GAN, cycleGAN. Cycle GANS are highly used for image-to-image translation where images of domain X are converted into images of domain Y. Further, Face swap GAN (FSGAN) which swaps the face of one person with another using facial landmarks of face. It has three major components i.e. reenactment generator, segmentation CNN and inpainting network. StarGAN is basically works on image translation framework for different domains. In this the style encoder is used to extract the required style from input and the mapping network extracts the style code from the latent vector and the corresponding domain given as input. Now the generator by using the mapping network or the style encoder generates the output image corresponding to input image. Another methods are STGAN and face swap-GAN \citep{patel2023deepfake}. 

\cite{melnik2024face} have conducted an in-depth analysis of face generation methods utilizing StyleGANs. The study discusses various StyleGAN variants, ranging from Progressive GAN to StyleGAN3, and explores their underlying latent representations, training metrics such as losses, and datasets. In addition, it delves into the inversion process, which involves converting an image into StyleGAN latent codes. The primary approaches highlighted include gradient-based optimization, which minimizes the loss between real and generated images to refine latent vectors, and training an encoder model on sample images to map them into latent space. A further technique discussed is fine-tuning the generator for specific tasks.

Similarly, \cite{cheng2024refining} address the limitations of existing GAN-based face generation methods, particularly the issues of domain irrelevance and insufficient representation of facial detail. Their proposed enhancement includes refining the generator architecture by integrating attention mechanisms and adaptive residual blocks to extract more nuanced facial features. A convolutional attention module is introduced after the residual block to emphasize important features in both the channel and the spatial dimensions, overcoming the original model’s inability to focus on key facial regions. Using the CelebA dataset that contains 202,599 face images with attribute labels, their algorithm demonstrates an improved ability to generate faces across various age groups and genders. The output images were visually and quantitatively evaluated, and the algorithm achieved a 20. 06\% and 14. 63\% improvement in the PSNR and SSIM values, respectively. Furthermore, \cite{khalid2021fakeavceleb} describe the Celeb-DF dataset, which uses 500 real YouTube videos of celebrities to generate deepfake videos using Faceswap methods. This dataset is widely used for evaluating the performance of deepfake generation and detection algorithms.

\cite{felouat2024ekyc} presented the eKYC-DF dataset, a pioneering resource designed to tackle identity fraud challenges in eKYC systems. This data set comprises real and synthetic facial videos, serving as a critical tool to advance and evaluate eKYC systems, especially in deep-fake detection and facial recognition. The paper outlines five significant contributions. First, it provides a comprehensive analysis of critical concepts and methods relevant to commercial eKYC solutions, deepfake generation and detection, commonly used datasets in deepfake detection, and the vulnerabilities of eKYC systems to deepfake-based spoofing. Second, the eKYC-DF dataset is significantly larger than most datasets used in deep learning, offering better support for pattern recognition and model generalization. Third, the dataset’s diversity, encompassing a wide range of ages, genders, and ethnicities, facilitates robust deep learning model training, minimizes biases, and improves prediction accuracy across different demographics. 
Finally, the paper sets a benchmark for assessing the effectiveness of the dataset in deepfake detection and facial matching, thereby enhancing its value in strengthening the security and reliability of eKYC systems.

\cite{tolosana2020deepfakes} provided a comprehensive review of face manipulation techniques in deepfake generation and detection, classifying them into four primary categories: complete face synthesis, identity swapping, attribute manipulation, and expression swapping. Entire face synthesis involves the use of advanced models like GANs and StyleGAN to generate entirely non-existent, high-quality face images that exhibit remarkable realism. The identity swap replaces one person's face with another using techniques such as face swapping. Attribute manipulation focuses on altering specific facial features, such as hair or skin color, gender, age, or adding accessories like glasses, often employing StarGAN-based approaches. Expression swapping, although similar to identity swapping, focuses on altering a person’s facial expressions while preserving their overall identity. These categories illustrate the diverse techniques used in facial manipulation within deep-fake technology.

\subsection{Deepfake Generation of Audio}

Synthetic audio data generation is primarily achieved through two approaches: text-to-speech (TTS) and voice conversion. Most synthetic speech generation methods utilize an encoder-decoder architecture, making GAN-based models well suited for this purpose. Examples of TTS models include Char2Wav, which employs a bidirectional RNN as the encoder and a conditional extension of SampleRNN as the decoder. Another notable model is WaveNet, which operates as a sequence-to-sequence encoder-decoder network. The first component of WaveNet predicts spectrograms using an encoder to process character sequences into intermediate features, which are then decoded into spectrograms. The second component, a modified version of WaveNet, generates time-domain waveform samples based on the input of the spectrogram from the first stage.

Other advanced TTS methods include WaveGlow, MelNet, Deep Voice 3, HiFi-GAN, and MelGAN, all of which leverage GANs or similar architectures for high-quality audio generation. Furthermore, voice impersonation represents another form of manipulation that goes beyond voice conversion by mimicking not only the target speaker’s speech and signal qualities but also their unique speaking style. For such tasks, GANs have been used for style transfer, with DiscoGAN being a notable example \citep{patel2023deepfake}.

The most widely used pre-trained models for deep audio generation include Tacotron 2, WaveNet, FastSpeech 2, VALL-E, WaveGlow, and TalkNet. \cite{wang2017tacotron} introduced Tacotron, a sequence-to-sequence model that synthesizes speech directly from input text, designed to enhance the vanilla seq2seq architecture. Process raw input characters and generate Mel spectrograms using an attention mechanism integrated into the seq2seq model.

Building on this, Tacotron 2 combines a neural network-based architecture with RNN and WaveNet to synthesize speech. The RNN model predicts Mel spectrogram sequences from input text using a sequence-to-sequence feature prediction network, while a modified version of WaveNet generates time-domain waveform samples from the predicted mel spectrograms. To evaluate Tacotron 2, subjective human evaluations, similar to Amazon’s Mechanical Turk, were performed, using the Mean Opinion Score (MOS) to assess the quality of the generated speech \citep{shen2018natural}.

\citep{wu2017asvspoof} highlights the ASVspoof challenges as pivotal in driving advancements in speaker verification research. The inaugural dataset, ASVspoof 2015, featured audio samples from 106 speakers, including both genuine and spoofed samples created using 10 distinct spoofing algorithms. Additionally, \citep{nautsch2021asvspoof} reports that the ASVspoof 2019 challenge broadened the scope by incorporating four types of spoofed audio data: text-to-speech synthesis, voice conversion, audio replay attacks, and advanced spoofing techniques. This expansion provided a comprehensive dataset for evaluating and enhancing speaker verification systems.

\subsection{Deepfake Generation of Audio-videos}

Previous studies have concentrated on generating deepfake videos, often overlooking the audio component. Many available datasets either feature only real audio or lack audio entirely. The FakeAVCeleb dataset was introduced to fill this gap, providing an audio-video deepfake dataset that includes both deepfake videos and synthesized, lip-synced fake audio. Created using generative models and real YouTube videos of celebrities, it offers a multimodal resource encompassing diverse ethnic backgrounds. The dataset is designed to aid in the development of multi-modal deepfake detectors \citep{khalid2021fakeavceleb}.

The DFDC dataset was developed as part of the Deepfake Detection Challenge (DFDC) in collaboration with researchers and organizations including Amazon Web Services, Facebook, and Microsoft \citep{dolhansky2020deepfake}. 
The DFDC dataset includes deepfake videos, synthesized cloned audio, and in some instances, both. These videos were created using eight distinct synthesis methods and recorded in diverse environmental conditions. However, a major limitation of the dataset is the absence of detailed labeling, especially regarding the alignment between audio and video. \cite{khalid2021fakeavceleb} introduced FakeAVCeleb, a novel multimodal deepfake detection dataset designed to address this limitation by providing synchronized labeling of deepfake videos and their corresponding synthesized cloned audio.

Fake audio generation involves integrating lip-syncing techniques with deepfake videos. Facial reenactment methods are employed to synchronize the synthesized audio with the target speaker, producing a deepfake video accompanied by corresponding fake audio \citep{khalid2021fakeavceleb}. The DFDC challenge concentrated on identifying whether the audio or video in a deepfake was fake. In comparison, the FakeAVCeleb dataset expands on this by addressing additional challenges, including varying environmental conditions, greater diversity, detailed audio-video labeling, and representation of individuals from different ethnicities, age groups, and genders. To create these deepfakes, a cloned voice of the target speaker is synthesized and synchronized with the video through facial reenactment techniques. Face-swapping tasks are performed using tools such as FaceSwap and FaceSwap GAN (FSGAN), while Wav2Lip is employed for facial reenactment based on source audio. For voice cloning, the Real-Time Voice Cloning (RTVC) tool is used. Additionally, synthesis processes are customized separately for each gender to enhance realism \citep{khalid2021fakeavceleb}.
\subsection{Deepfake Generation for Indian/Multilingual/Low Resource Languages}

\cite{hou2024polyglotfake} have designed multilingual dataset for deepfakes i.e. PolyGlotFake. This dataset contains content in seven languages and designed using all the Text-to-Speech, voice cloning, and lip-sync techniques. The PolyGlotFake dataset comprises a total of 15238 videos, including 766 real videos and 14472 fake videos. The average duration of each video is 11.79 seconds, with a resolution of $1280 \times 720$. Apart from this we have many recent works in multilingual deepfakes such as~\citep{shelar2022deepfakes, seong2021multilingual}\\

Authors have proposed audio based deepfake dataset for Urdu language using text to speech synthesis models like Tacotron and VITS TTS. The proposed dataset have been evaluated qualitative and quantitative using human evaluator,  Equal Error Rate(EER), t-SNE plotting and comparing L2 norms. It uses 495 text sentences from Urdu news sources \citep{munir2024deepfake}.

\section{Methodology}

\subsection{Real Data Collection}
The HAV-DF dataset comprises 508 high-quality videos, including 200 authentic and 308 synthetically manipulated samples. All source videos were carefully curated from YouTube, adhering to strict selection criteria to ensure consistency and quality across the dataset. The selection of the videos were based on the frontal view of the subject, single person per video, subject in a straight and upright position, clear Hindi audio, Unobstructed centered face and minimal use/absence of accessories (e.g., hats, glasses, masks) that might occlude facial features. Further, the dataset includes a mix of videos featuring both celebrities and ordinary individuals, ensuring a wide range of facial characteristics and speaking styles. The topics covered in the videos span news segments, technology discussions, and podcast excerpts, providing a diverse representation of Hindi speech patterns and content. The dataset is nearly equally balanced between male and female subjects.

\subsection{Proposed Architecture for HAV-DF}

\begin{figure}[!h]
    \centering
\includegraphics[width=\linewidth]{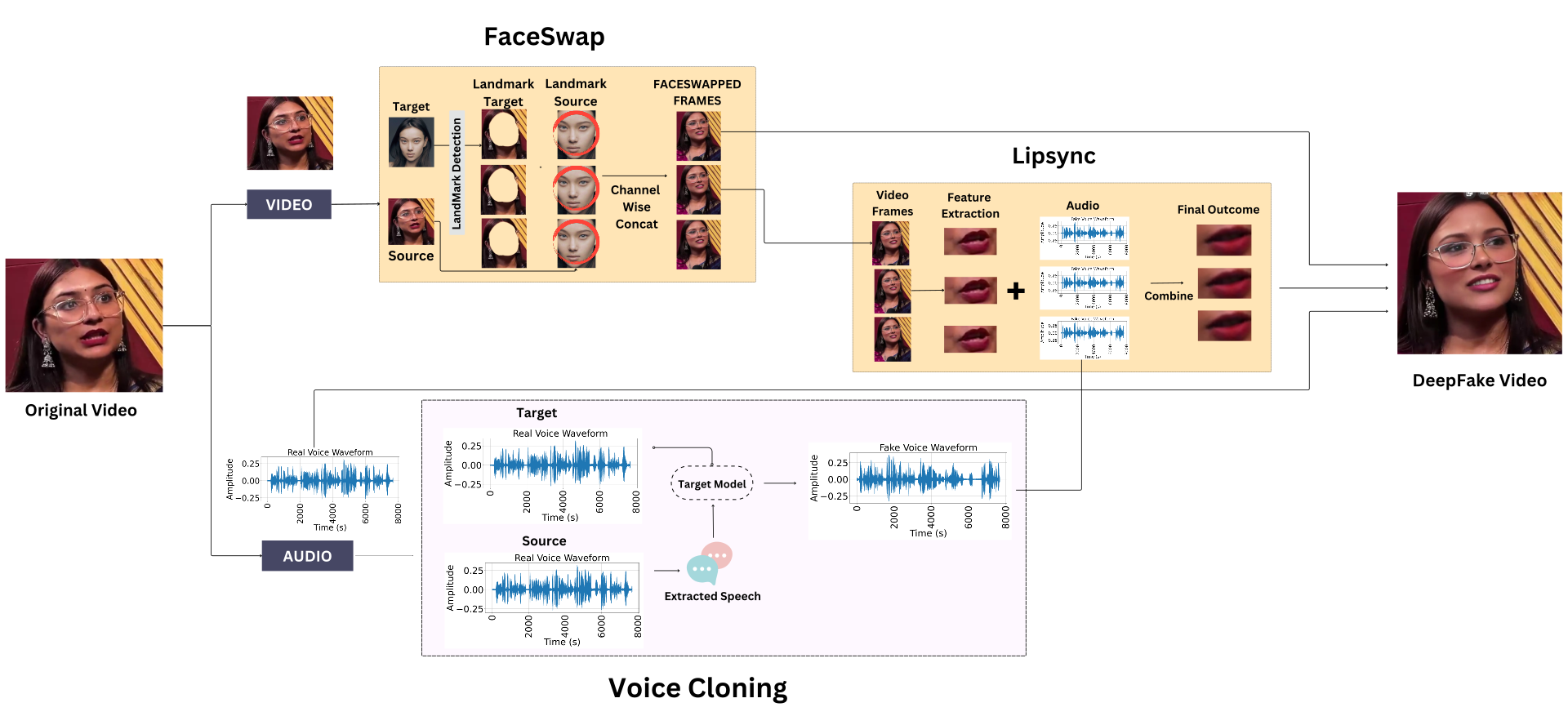} 
\caption{Demonstration of HAV-DF dataset generation pipeline.}
\label{fig:1}
\end{figure}

Figure~\ref{fig:1} demonstrate the comprehensive methodology employed to create  novel Hindi-language audio-video deepfake dataset, \textit{Hindi audio-video-Deepfake (HAV-DF)}. The approach integrates three sophisticated AI techniques to produce highly convincing deepfakes.
\begin{itemize}
    \item \textbf{Face swap:} Replace the original faces in source videos with target faces, using advanced landmark detection and face swapping algorithms across multiple frames. For this, FSGAN~\citep{nirkin2019fsgan}, FaceSwap~\citep{liu2023deepfacelab}, and DeepFaceLab~\citep{perov2020deepfacelab} tools have been used.
    
    \item  \textbf{Lip sync:} Lip features has been extracted from the face-swapped frames and combined with audio waveforms to ensure realistic lip movements that match the speech. For this, ReTalking method has been used~\citep{cheng2022videoretalking}. An example of manipulated audio is shown in Fig~\ref{fig:pitch_contour}.
    
    \item  \textbf{Voice Cloning:} For audio manipulation, this process analyzes the original Hindi speech alongside a target voice model to generate a synthetic voice that maintains the content of the original speech but mimics the characteristics of the target speaker. For this, RVC Model has been used~\citep{kambali2023real}.
\end{itemize}

The final output of the proposed pipeline is a diverse set of deepfake videos that form the HAV-DF dataset. Each video in the dataset combines face-swapped and lip-synced visuals with cloned voices, resulting in highly realistic synthetic media where facial identity, lip movements, and voice have all been artificially manipulated.

\subsection{Fake Audio}
\begin{figure}[!h]
    \centering
\includegraphics[width=0.45\linewidth]{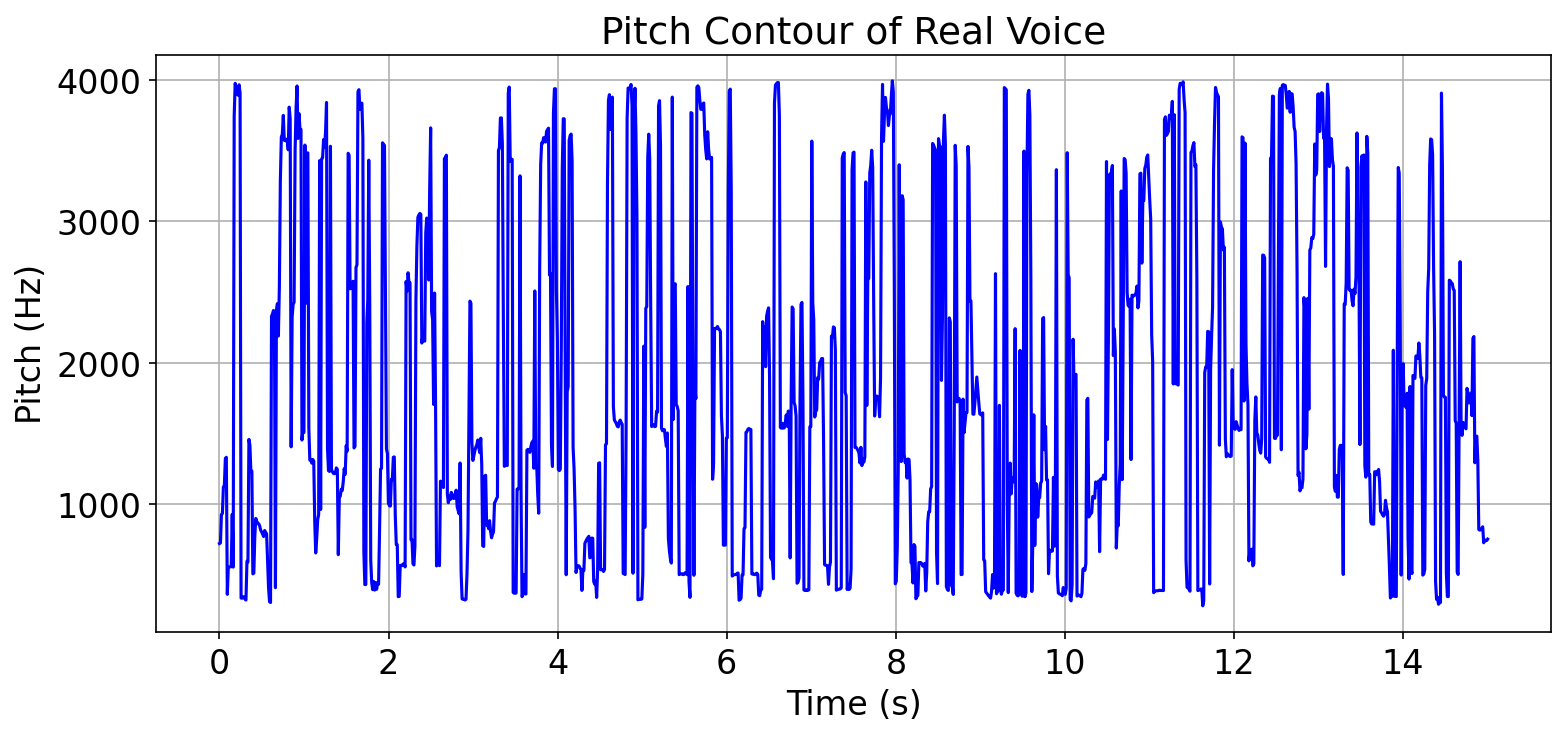} 
\includegraphics[width=0.45\linewidth]{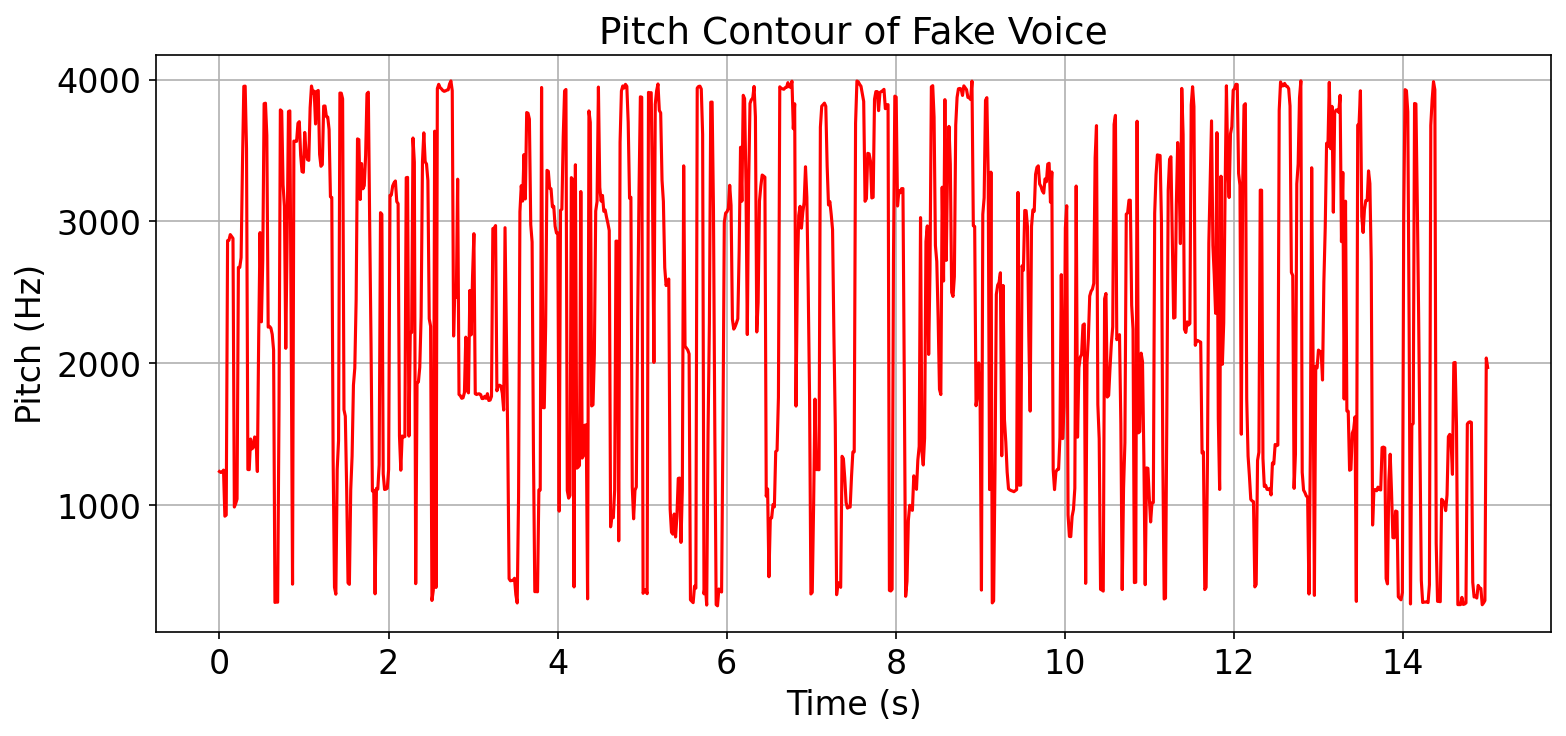} 
\caption{Comparison of pitch contours between real and fake voice samples from the HAV-DF dataset. }

\label{fig:pitch_contour}
\end{figure}

Figure~\ref{fig:pitch_contour} present pitch contours of audio samples, comparing a real voice recording (left) to a synthetic or deepfake voice (right) from the HAV-DF dataset. Pitch contours visualize the fundamental frequency of speech over time, providing insights into intonation patterns and voice characteristics.
Both figures show pitch variations over a 15-second duration, with the vertical axis representing pitch frequency (Hz) and the horizontal axis representing time(s). Figure~\ref{fig:pitch_contour}(left) (Real Voice) displays natural pitch variations of human speech. The contour shows a wide range of frequencies, from about 500 Hz to 4000 Hz, with smooth transitions between different pitch levels. There are clear patterns of rising and falling intonation, likely corresponding to sentence structures and emotional expressions in natural speech.
Figure~\ref{fig:pitch_contour}(right) (Fake Voice) also presents a complex pitch contour, attempting to mimic natural speech patterns. However, there are subtle differences compared to the real voice. The fake voice seems to have more abrupt changes in pitch and possibly a slightly narrower overall pitch range. The contour appears somewhat more erratic, with more frequent jumps between high and low frequencies.
These pitch contours highlight the sophistication of voice synthesis technology in the HAV-DF dataset, as the fake voice closely mimics many characteristics of natural speech. However, the differences in pitch variation patterns and smoothness of transitions could potentially be leveraged by deepfake detection algorithms. The comparison underscores the challenges in distinguishing between real and synthetic voices, especially in non-English languages like Hindi, and emphasizes the need for advanced acoustic analysis in deepfake detection research.

\subsection{Detailed Dataset Design Models}

\begin{figure}[!h]
    \centering
\includegraphics[width=0.85\linewidth]{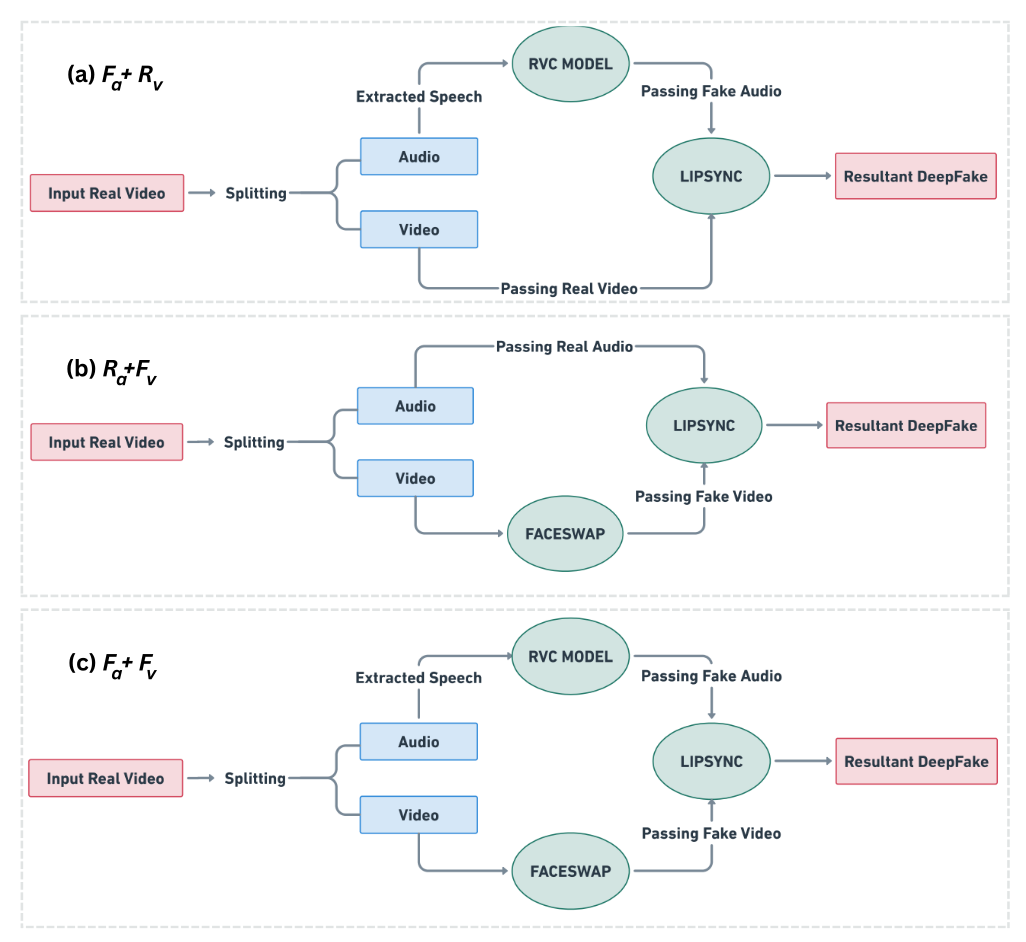} 

\caption{Architecture of the HAV-DF dataset. It consists of three sets: (\textit{upper}) $F_a+R_v$: fake audio annd real video, (\textit{middle}) $R_a+F_v$: real audio and fake video, and (\textit{lower}) $F_a+F_v$: fake audio and fake video.}

\label{fig:flowchart}
\end{figure}
Figure~\ref{fig:flowchart} illustrates the process of creating three types of deepfakes in the HAV-DF dataset. 

\begin{itemize}
    \item \textbf{Fake audio and real video ($F_a+R_v$):} The topmost flow chart, labeled as \textit{$F_a+R_v$}   explains the process to create a fake with fake audio and real video. Here, an input real video is split into audio and video components, and then the audio is passed through a realistic voice cloning (RVC) model to generate fake audio. This fake audio is then combined with the original real video using a Lipsync model to produce the final deepfake.
    
    \item \textbf{Real audio and fake video (\textit{$R_a+F_v$}): } The central flow chart, labeled \textit{$R_a+F_v$}, illustrates the process of generating a deep-fake with real audio and fake video. In this method, an input real video is separated into its audio and video components while retaining the original audio. The video component is processed using the FaceSwap technique to produce fake video content. Finally, real audio and fake video are synchronized using the Lipsync model to create the deepfake.

    \item \textbf{Fake audio and fake video (\textit{$F_a+F_v$}):} The lower flow chart, labeled \textit{$F_a+F_v$}, depicts the most intricate deepfake creation process, where both audio and video are manipulated. The input video is separated into its components, with the audio processed by the RVC model to generate fake audio. At the same time, the video undergoes FaceSwap to produce fake video content. These manipulated audio and video elements are then synchronized using the Lipsync model to create the final deepfake. This detailed diagram effectively showcases the advanced techniques used to create complex deepfakes, emphasizing the challenges involved in detecting such manipulations.
\end{itemize}

\subsubsection{Face Swap Process}
\begin{figure}[!ht]
    \centering
\includegraphics[width=0.65\linewidth]{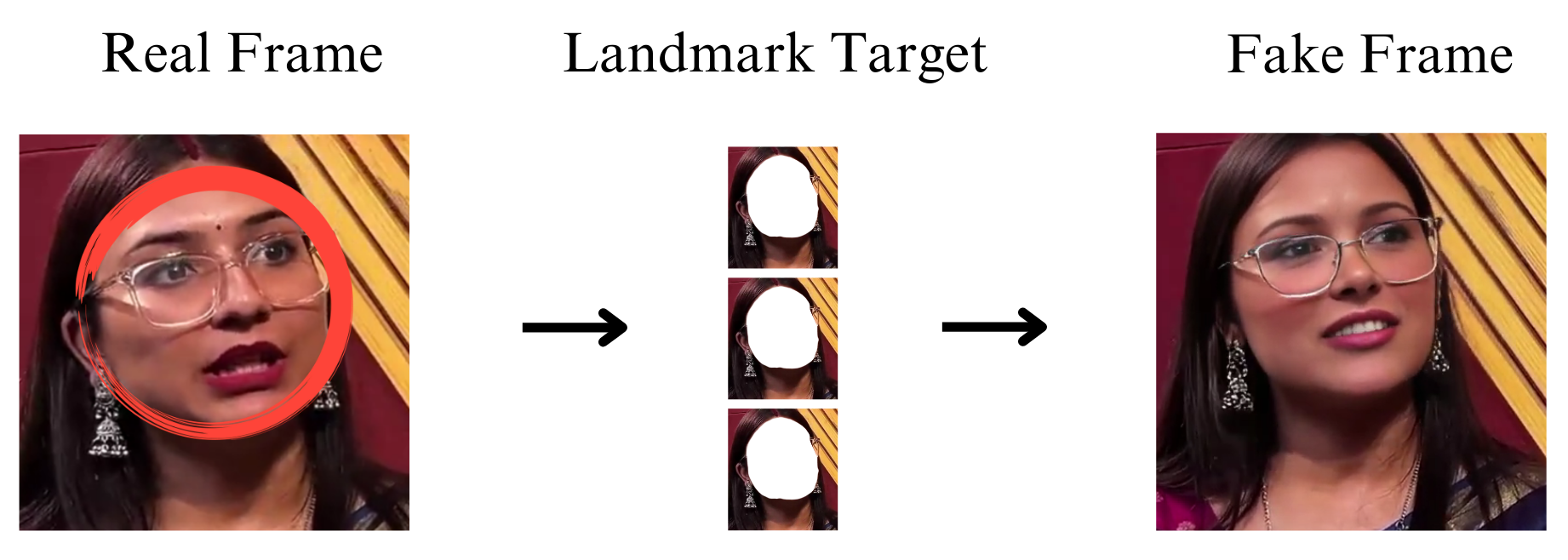} 
\caption{Demonstration of FaceSwap deepfake generation process.}
\label{fig:FaceSwap}
\end{figure}

Figure~\ref{fig:FaceSwap} illustrates the face-swapping process utilized in the creation of deepfake videos within the HAV-DF dataset. This process comprises three key stages: Real Frame, Landmark Target, and Fake Frame.
\begin{itemize}
    \item \textit{Real Frame:} The initial stage presents the original video frame, featuring a woman with distinctive characteristics such as long dark hair, large glasses, and ornate earrings. A red circular overlay emphasizes the facial region targeted for manipulation.

    \item \textit{Landmark Target:} The intermediate stage displays a series of simplified facial outlines, representing key facial landmarks. These landmarks serve as crucial reference points for the face-swapping algorithm, ensuring proper alignment and integration of the new facial features.

    \item \textit{Fake Frame:} The final stage showcases the resultant frame post-face-swapping. While maintaining the original frame's contextual elements such as hair, background, and accessories, the core facial features exhibit subtle yet significant alterations. These changes include a modified face shape, a more neutral expression, and slight variations in the glasses design.
\end{itemize}


\subsubsection{Lip-Sync Process}
\begin{figure}[!ht]
    \centering
\includegraphics[width=0.65\linewidth]{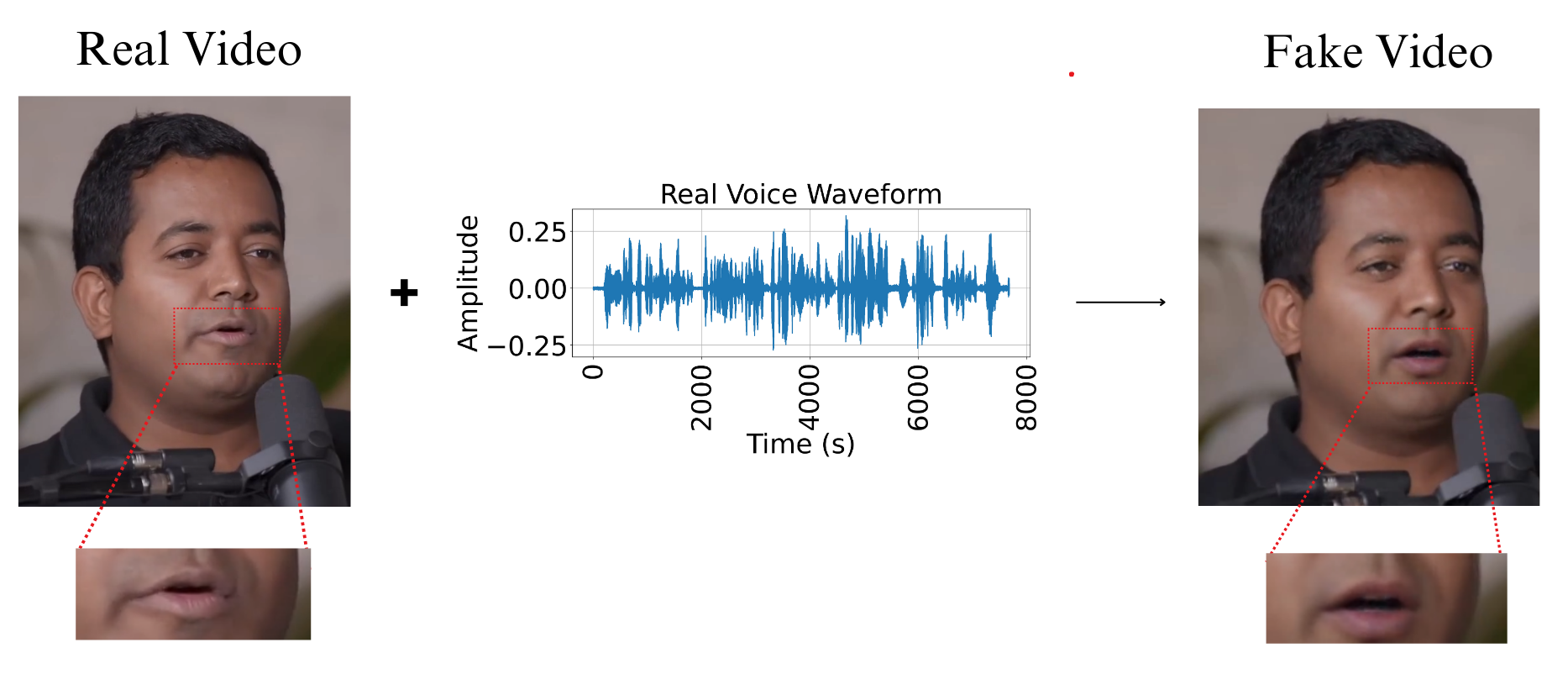} 
\caption{Demonstration of Lip Syncing deepfake generation process.}
\label{fig:LipSync}
\end{figure}
Figure~\ref{fig:LipSync} illustrates the creation of a lip-synced deepfake video using real audio and a manipulated video. On the left, the `Real Video' shows a man speaking, with a zoomed-in inset of his mouth area. In the center, a `Real Voice Waveform' graph represents the audio signal of the genuine speech, showing amplitude variations over time. On the right, the `Fake Video' result appears nearly identical to the real video, but with subtle differences in the mouth area, as shown in the zoomed-in inset. An arrow points from the combination of real video and voice waveform to the fake video, indicating the synthesis process. This demonstrates how lip-syncing technology can create a convincing video where lip movements are manipulated to match an existing audio track while maintaining the speaker's overall appearance and identity, highlighting the sophisticated manipulation possible in modern deepfake technology.

\subsubsection{Voice Cloning}
\begin{figure}[!h]
    \centering
\includegraphics[width=0.45\linewidth]{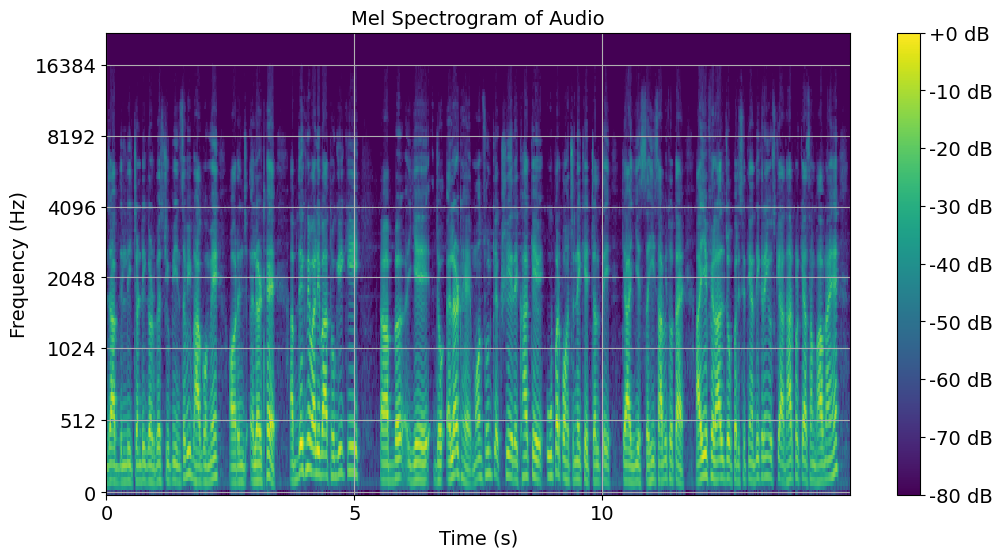} 
\includegraphics[width=0.45\linewidth]{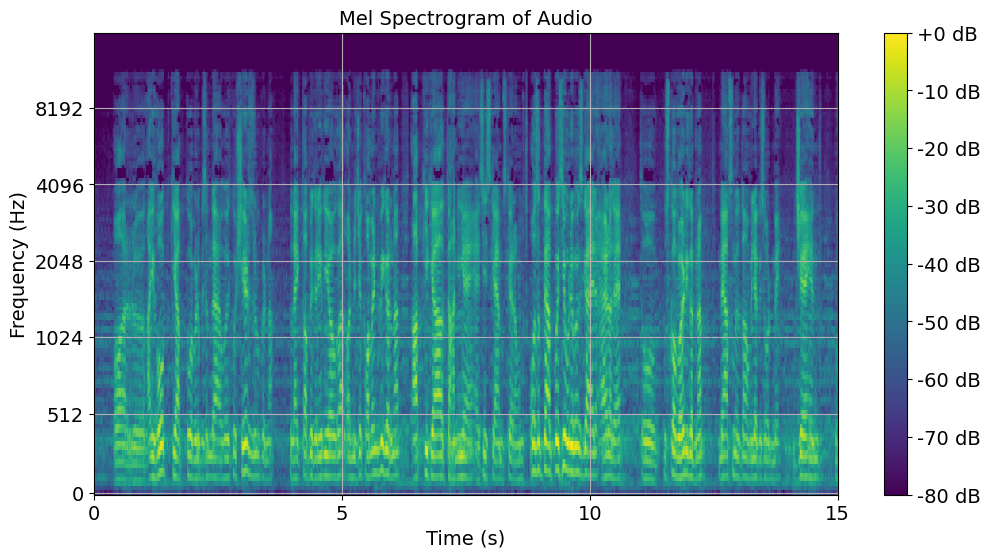} 
\caption{Mel spectrograms of real and synthetic audio samples from the HAV-DF dataset.}

\label{fig:voice_clone}
\end{figure}
Figure~\ref{fig:voice_clone} shows Mel spectrograms of audio samples, likely comparing a real audio recording (left) to a synthetic or deepfake audio (right) from HAV-DF dataset. Mel spectrograms visualize the frequency content of audio signals over time, with frequency on the vertical axis, time on the horizontal axis, and color intensity representing the energy at each time-frequency point.
Left figure displays a spectrogram with a wider frequency range (up to 16384 Hz) and covers about 15 seconds of audio. It shows clear harmonic structures and formant patterns typical of natural speech, with strong energy concentrations in the lower frequencies (below 4000 Hz) and distinct temporal variations. On the other hannd, right figure covers a slightly longer duration (15 seconds) but with a lower maximum frequency (8192 Hz). While it also shows speech-like patterns, there are subtle differences in the energy distribution and temporal structure compared to real spectrogram (on left). The synthetic audio appears to have less defined harmonic structures in the higher frequencies and some artifacts or inconsistencies in the spectral patterns.
These spectrograms highlight the challenges in detecting audio deepfakes, as the synthetic audio (on right) closely mimics the general spectral characteristics of natural speech (on left). However, the differences in spectral detail and energy distribution, particularly in higher frequencies, could potentially be used by deepfake detection algorithms to distinguish between real and synthetic audio in the HAV-DF dataset. The comparison underscores the sophistication of modern voice synthesis techniques and the need for advanced analysis methods in deepfake detection.
\subsubsection{Challenges Faced}
The development of the HAV-DF dataset presented several unique challenges, reflecting the complexities of creating a comprehensive Hindi-language deepfake dataset:
\begin{itemize}
    \item Locating high-quality Hindi videos meeting all selection criteria proved time-consuming.
    \item Ensuring a balance between celebrity and non-celebrity content while maintaining diversity was challenging.
    \item Hindi's rich phonetic structure and diverse accents complicated the voice cloning process.
    \item Lip-syncing algorithms required fine-tuning to accurately match Hindi phonemes.
    \item Preserving culturally specific gestures and expressions during face swapping was crucial but technically demanding.
    \item Adapting existing deepfake algorithms, primarily optimized for English content, to Hindi speech patterns required significant effort.
    \item Ensuring uniformity in video quality, duration, and manipulation effectiveness across the dataset required extensive manual review and adjustments.
    \item Generating high-quality deepfakes for 350 videos demanded substantial computational power and time.
\end{itemize}

\section{Experimental}

\subsection{HAV-DF Dataset Description}

\begin{figure}[!ht]
    \centering
\includegraphics[width=0.85\linewidth]{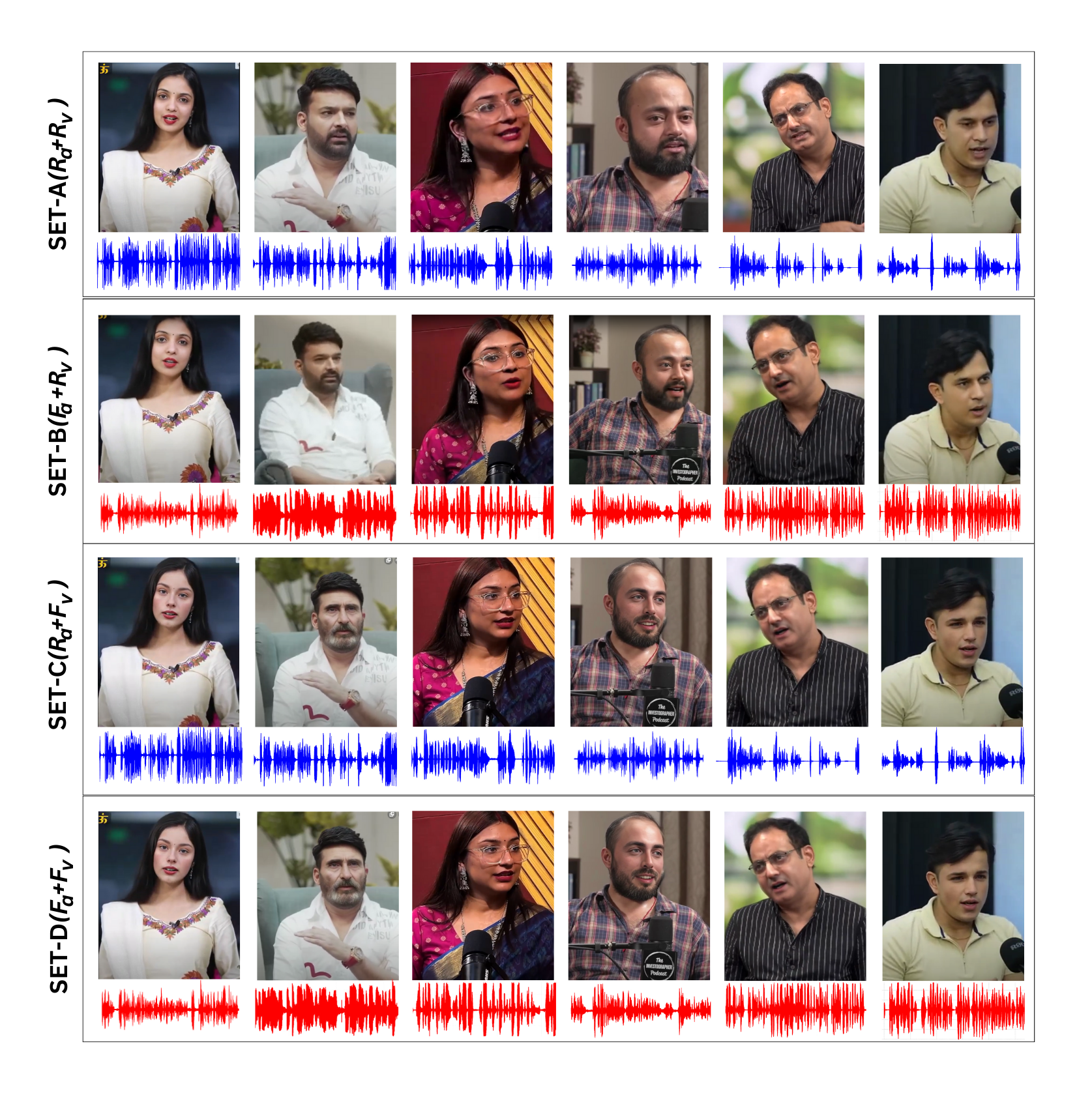} 
\caption{Samples from the HAV-DF dataset. It consists of four sets: (upper) $R_a+R_v$: real audio and real video, (upper middle) $F_a+R_v$: fake audio and real video, (\textit{lower middle}) $R_a+F_v$: real audio and fake video and (\textit{lower})$F_a+F_v$: fake audio and fake video.}
\label{fig:dataset}
\end{figure}

Figure~\ref{fig:dataset} illustrates the various components of the HAV-DF dataset, showing different combinations of real and fake audio and video content. The image is divided into four rows, each representing a distinct set of audio-video combinations. Each row contains six individual portraits of different individuals, presumably speaking in Hindi, accompanied by corresponding audio waveforms beneath each portrait.

\subsubsection{Real Dataset ($R_a+R_v$)}
We sampled 200 videos with diverse ethnicity, gender, and age from YouTube. The duration of each video is 15 seconds with resolution of $1080$ p ($1920 \times1080$ pixels). Each video frame rate is 30 frames per second. The video selection criteria includes frontal view of the subject, single person per video, subject in a straight and upright position, and clear Hindi audio. (see Fig~\ref{fig:dataset}(upper)).

\subsubsection{Deepfake Dataset}

A total of 308 deepfake Hindi videos were generated by maintaining the ethnicity, gender, and age of each subject in the video. The duration of each video is 15 seconds with a resolution of $1080$ p ($1920 \times1080$ pixels). Each video frame rate is 30 frames per second. The deepfake dataset contains a total of 78 \textit{Fake audio and real video ($F_a+R_v$)}, 130 of \textit{Real audio and fake video ($R_a+F_v$)}, and 100 of \textit{Fake audio and fake video }($F_a+F_v$). Fig~\ref{fig:dataset} shows the samples of audio-video from videos frame. The upper row, labeled SET-A ($R_a+R_v$), probably represents real audio with real video, serving as the authentic baseline content. The second row, SET-B ($F_a+R_v$) appears to show fake or generated audio (indicated by red waveforms) paired with real video. The third row, SET-C ($R_a+F_v$) depicts real audio with fake or manipulated video. Finally, the lower row, SET-D ($F_a+F_v$) represents both fake audio and fake video content. 


\section{Evaluation Metrics}

Here, we compare the HAV-DF dataset with other existing deepfake datasets and will perform the qualitative and quantitative analysis.
\subsection{Qualitative Analysis}

\begin{figure}[!ht]
    \centering
\includegraphics[width=0.75\linewidth]{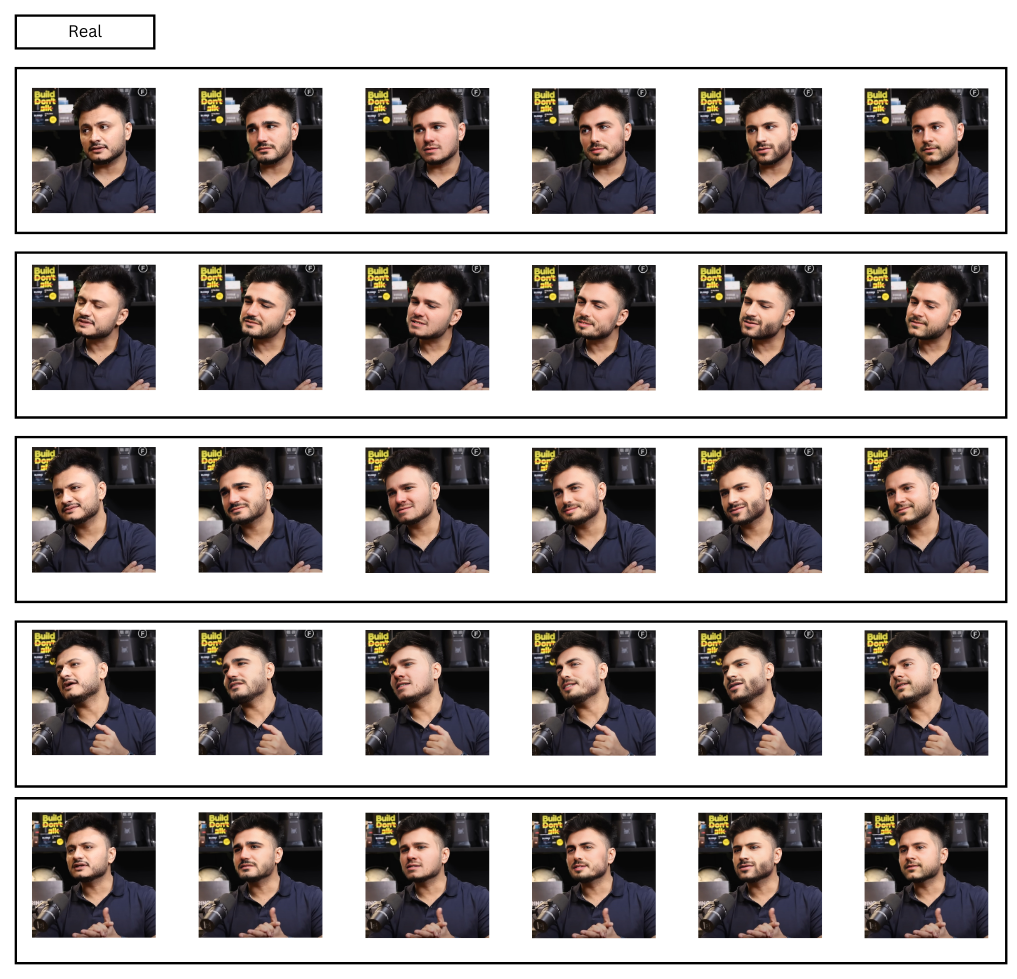} 
\caption{Face swapping results and quality of HAV-DF dataset. Example target images showing diversity for pose, facial expression, and illumination. }
\label{fig:DQ1}
\end{figure}

\begin{figure}[!ht]
    \centering
\includegraphics[width=0.75\linewidth]{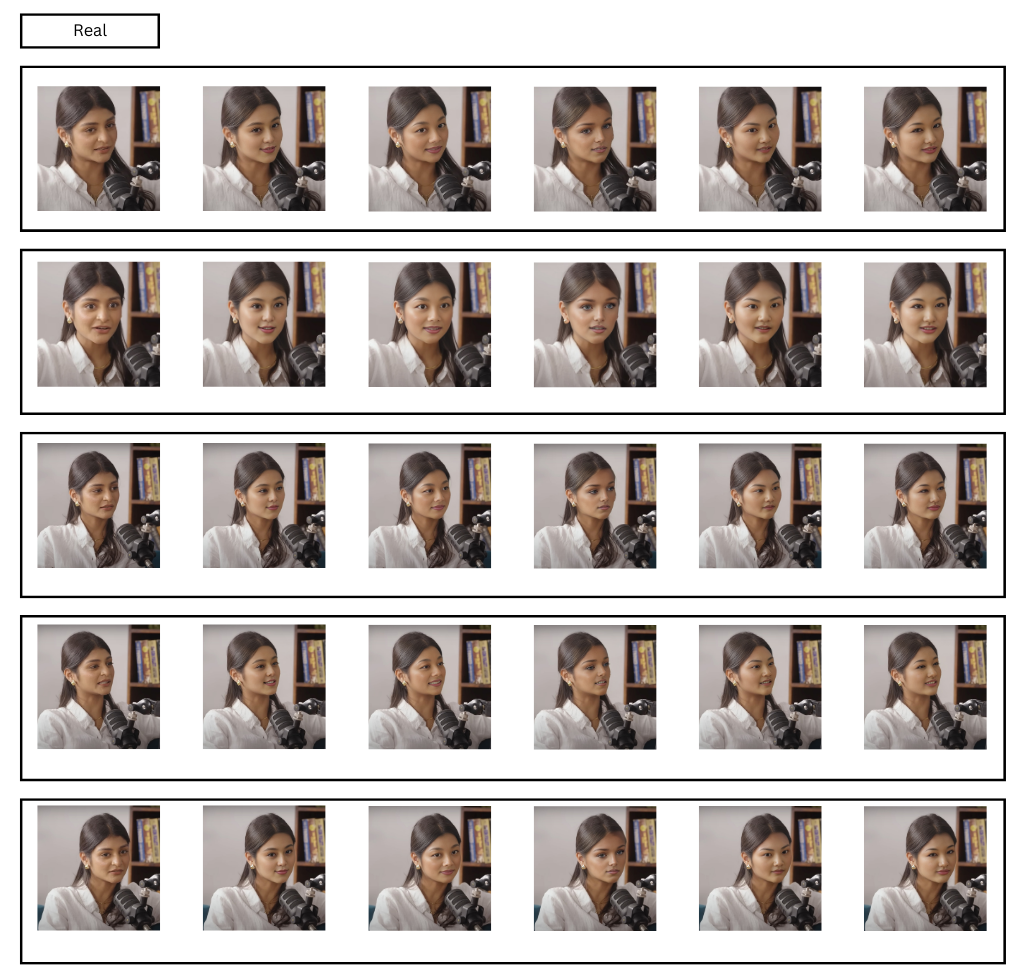} 
\caption{Face swapping results and quality of HAV-DF dataset. Example target images showing diversity for pose, facial expression, and illumination. }
\label{fig:DQ2}
\end{figure}

Figures~\ref{fig:DQ1} and \ref{fig:DQ2} show the quality of video frames generated by the face wapping method for the HAV-DF dataset. The data set shows the diversity for age, gender and ethnicity, along with the diversity in pose, facial expression, and illumination. In figure~\ref{fig:DQ1}, each row contains a sequence of frames, where subtle differences can be observed in facial expressions, head orientation, or other facial features. The first column is labeled ``Real'', serving as a reference for comparison with subsequent frames. Generated frames are evaluated for their closeness to the real frame in terms of facial features, skin texture and lighting, natural expressions, and lip-sync accuracy. Common issues like blurring, distortions, or mismatches in facial expressions between frames are checked.
\subsection{Quantitative Analysis}
In quantitative analysis, the HAV-DF dataset has been compared with other video only and audio-video datasets across various deepfake detection methods. The dataset is divided into training and testing sets as 408 videos for training and 100 videos for testing.
 Table~\ref{table:DataInfo1} provides the detailed information of various deepfake datasets: UADFV~\citep{xie2020deepfake}, 
Deepfake TIMIT~\citep{korshunov2018deepfakes}, 
FF++~\citep{rossler2019faceforensics++}, DFDC~\citep{dolhansky2020deepfake}, 
Celeb-DF~\citep{li2020celeb},
Deeper Forensics-1.0~\citep{jiang2020deeperforensics},
KoDF~\citep{kwon2021kodf}, 
DF-Platter~\citep{narayan2023df},  
DFFD~\citep{dang2020detection},
WildDeepfake~\citep{zi2020wilddeepfake}, 
FakeAVCeleb~\citep{khalid2021fakeavceleb}, and ForgreyNet~\citep{he2021forgerynet}.
Similarly, Table~\ref{table:DataInfo2} provides a comparative analysis of various deepfake generation methods and models.

\begin{table}[!ht]
\caption{Table summarizing various deepfake datasets including their launch dates, generation, media type, real and fake content counts, sources of real and fake media, and usage rates, highlighting developments from 2018 to 2024.}
\begin{tabular}{|l|c|c|c|l|c|c|c|}
\hline
\textbf{Dataset Name}  & \textbf{\begin{tabular}[c]{@{}c@{}}Launch\\ Date\end{tabular}} & \textbf{Gen} & \textbf{\begin{tabular}[c]{@{}c@{}}Images/\\ Videos\end{tabular}} & \textbf{\begin{tabular}[c]{@{}l@{}}Real/\\ Fake\end{tabular}} & \textbf{\begin{tabular}[c]{@{}c@{}}Source of \\ Real\end{tabular}}      & \textbf{\begin{tabular}[c]{@{}c@{}}Source of \\ Fake\end{tabular}} & \textbf{\begin{tabular}[c]{@{}c@{}}Usage \\ Rate\end{tabular}} \\ \hline
\textbf{UADFV} & 2018  & 1            & Videos   & 49/49   & YouTube  & Artificial       & 5\%                                                            \\ \hline
\textbf{\begin{tabular}[c]{@{}l@{}}Deepfake\\ TIMIT\end{tabular}}      & 2018     & 1            & Videos    & \begin{tabular}[c]{@{}l@{}}320/\\ 640\end{tabular}            & Vid-TIMIT & Artificial & 11\% \\ \hline
\textbf{FF++}                                                           & 2019                                                           & 1            & Videos                                                            & \begin{tabular}[c]{@{}l@{}}1000/\\ 4000\end{tabular}          & YouTube                                                                 & Artificial                                                         & 51\%                                                           \\ \hline
\textbf{DFDC}                                                           & 2020                                                           & 2            & Videos                                                            & \begin{tabular}[c]{@{}l@{}}23,654/\\ 104,500\end{tabular}     & \begin{tabular}[c]{@{}c@{}}Volunteer \\ Actors\end{tabular}             & Artificial                                                         & 20\%                                                           \\ \hline
\textbf{Celeb-DF}                                                       & 2020                                                           & 2            & Videos                                                            & \begin{tabular}[c]{@{}l@{}}590/\\ 5639\end{tabular}           & YouTube                                                                 & Artificial                                                         & 34\%                                                           \\ \hline
\textbf{\begin{tabular}[c]{@{}l@{}}Deeper\\ Forensics-1.0\end{tabular}} & 2020                                                           & 3            & Videos                                                            & \begin{tabular}[c]{@{}l@{}}1,000/\\ 60,000\end{tabular}       & \begin{tabular}[c]{@{}c@{}}Controlled \\ studio footage\end{tabular}    & Artificial                                                         & N/A                                                            \\ \hline
\textbf{KODF}                                                           & 2021                                                           & 2            & Videos                                                            & \begin{tabular}[c]{@{}l@{}}1,000/\\ 10,000\end{tabular}       & \begin{tabular}[c]{@{}c@{}}YouTube \\ (Korean Celebrities)\end{tabular} & Artificial                                                         & N/A                                                            \\ \hline
\textbf{DF-Platter}                                                     & 2023                                                           & 2            & Videos                                                            & \begin{tabular}[c]{@{}l@{}}764/\\ 132,496\end{tabular}        & YouTube                                                                 & Artificial                                                         & N/A                                                            \\ \hline
\textbf{DFFD}                                                           & 2020                                                           & 2            & \begin{tabular}[c]{@{}c@{}}Images/\\ Videos\end{tabular}          & \begin{tabular}[c]{@{}l@{}}58,703/\\ 240,336\end{tabular}     & \begin{tabular}[c]{@{}c@{}}Various\\ internet sources\end{tabular}      & Artificial                                                         & 1\%                                                            \\ \hline
\textbf{WildDeepfake}                                                   & 2020                                                           & 2            & Videos                                                            & \begin{tabular}[c]{@{}l@{}}3,805/\\ 3,509\end{tabular}        & Internet                                                                & Internet                                                           & 1\%  \\ \hline
\textbf{FakeAVCeleb}  & 2021 & 2            & Videos    & \begin{tabular}[c]{@{}l@{}}500+ / \\ 20,000+\end{tabular}     & \begin{tabular}[c]{@{}c@{}}YouTube\\ (VoxCeleb)\end{tabular}            & Artificial                                                         & N/A                                                            \\ \hline \textbf{ForgeryNet}                                                     & 2021                                                           & 3            & \begin{tabular}[c]{@{}c@{}}Images/\\ Videos\end{tabular}          & \begin{tabular}[c]{@{}l@{}}2.9M/\\ 9.3M\end{tabular}          & Various sources                                                         & Artificial                                                         & N/A \\ \hline
\textbf{HAV-DF}                                                         & 2024                                                           & 1            & Videos                                                            & \begin{tabular}[c]{@{}l@{}}100/\\ 400\end{tabular}            & YouTube                                                                 & Artificial                                                         & N/A                                                            \\ \hline
\end{tabular}
\label{table:DataInfo1}
\end{table}
\begin{table}[!h]
\caption{Table detailing the manipulation methods and models used in various deepfake datasets, categorizing the presence of lip-sync, face swap, and audio manipulation technologies.}
\begin{tabular}{|l|c|c|c|l|}
\hline
\textbf{Dataset Name}                                                   & \textbf{\begin{tabular}[c]{@{}c@{}}Video \\ Manipulation\\  (Lip-sync)\end{tabular}} & \textbf{Face Swap}                                                               & \textbf{\begin{tabular}[c]{@{}c@{}}Audio \\ Manipulation\end{tabular}}      & \textbf{\begin{tabular}[c]{@{}l@{}}Models Used \\ for Manipulation\end{tabular}}                                                      \\ \hline
\textbf{UADFV}                                                          & No                                                                                   & Yes                                                                              & No                                                                          & \begin{tabular}[c]{@{}l@{}}FaceSwap \\ (using autoencoders)\end{tabular}                                                              \\ \hline
\textbf{\begin{tabular}[c]{@{}l@{}}Deepfake \\ TIMIT\end{tabular}}      & No                                                                                   & Yes                                                                              & No                                                                          & GAN-based FaceSwap                                                                                                                    \\ \hline
\textbf{FF++}                                                           & \begin{tabular}[c]{@{}c@{}}Yes \\ (e.g., Face2Face)\end{tabular}                     & \begin{tabular}[c]{@{}c@{}}Yes \\ (FaceSwap, \\ DeepFakes)\end{tabular}          & \begin{tabular}[c]{@{}c@{}}Yes \\ (NeuralTextures)\end{tabular}             & \begin{tabular}[c]{@{}l@{}}Face2Face, FaceSwap, \\ DeepFakes, NeuralTextures\end{tabular}                                             \\ \hline
\textbf{DFDC}                                                           & \begin{tabular}[c]{@{}c@{}}Yes \\ (Neural Talking \\ Heads)\end{tabular}             & \begin{tabular}[c]{@{}c@{}}Yes \\ (DFAE, FSGAN, \\ MM/NN, StyleGAN)\end{tabular} & \begin{tabular}[c]{@{}c@{}}Yes \\ (TTS Skins)\end{tabular}                  & \begin{tabular}[c]{@{}l@{}}DFAE, MM/NN, Neural \\ Talking Heads, FSGAN, \\ StyleGAN, TTS Skins\end{tabular}                           \\ \hline
\textbf{Celeb-DF}                                                       & No                                                                                   & Yes                                                                              & No                                                                          & \begin{tabular}[c]{@{}l@{}}Improved DeepFake \\ synthesis (FaceSwap, \\ DeepFakes)\end{tabular}                                       \\ \hline
\textbf{\begin{tabular}[c]{@{}l@{}}Deeper\\ Forensics-1.0\end{tabular}} & No                                                                                   & Yes                                                                              & No                                                                          & \begin{tabular}[c]{@{}l@{}}DFAE, GAN-based\\ methods (StyleGAN)\end{tabular}                                                          \\ \hline
\textbf{KODF}                                                           & Yes (LipGAN)                                                                         & \begin{tabular}[c]{@{}c@{}}Yes \\ (DeepFaceLab, \\ FaceSwap)\end{tabular}        & \begin{tabular}[c]{@{}c@{}}Yes \\ (Voice Conversion \\ Models)\end{tabular} & \begin{tabular}[c]{@{}l@{}}DeepFaceLab, \\ FaceSwap, LipGAN\end{tabular}                                                              \\ \hline
\textbf{DF-Platter}                                                     & No                                                                                   & Yes                                                                              & No                                                                          & \begin{tabular}[c]{@{}l@{}}FSGAN, FaceSwap,\\ FaceShifter\end{tabular}                                                                \\ \hline
\textbf{DFFD}                                                           & No                                                                                   & Yes                                                                              & No                                                                          & \begin{tabular}[c]{@{}l@{}}FaceSwap, DeepFakes, \\ StyleGAN\end{tabular}                                                              \\ \hline
\textbf{WildDeepfake}                                                   & No                                                                                   & Yes                                                                              & No                                                                          & \begin{tabular}[c]{@{}l@{}}FaceSwap-GAN, \\ DeepFaceLab, \\ Faceswap, DeepFakes, \\ FSGAN, Others\end{tabular}                        \\ \hline
\textbf{FakeAVCeleb}                                                    & \begin{tabular}[c]{@{}c@{}}Yes \\ (Wav2Lip)\end{tabular}                             & \begin{tabular}[c]{@{}c@{}}Yes \\ (FSGAN, FaceSwap, \\ DeepFaceLab)\end{tabular} & Yes (RTVC)                                                                  & \begin{tabular}[c]{@{}l@{}}FSGAN, FaceSwap, \\ DeepFaceLab, \\ Wav2Lip, RTVC\end{tabular}                                             \\ \hline
\textbf{ForgeryNet}                                                     & Yes                                                                                  & Yes                                                                              & Yes                                                                         & \begin{tabular}[c]{@{}l@{}}Various GANs, FSGAN,\\  Audio-based Manipulation, \\ DeepFakes, FaceShifter,\\ NeuralTextures\end{tabular} \\ \hline
\textbf{HAV-DF}                                                         & Yes (ReTalking)                                                                      & \begin{tabular}[c]{@{}c@{}}Yes \\ (FSGAN, FaceSwap,\\ DeepFaceLab)\end{tabular}  & Yes (RVC Model)                                                             & \begin{tabular}[c]{@{}l@{}}FSGAN, FaceSwap,\\ FaceShifter, Retalking, \\ RVC Model\end{tabular}                                       \\ \hline
\end{tabular}
\label{table:DataInfo2}
\end{table}
\begin{figure}[!ht]
    \centering
\includegraphics[width=0.65\linewidth]{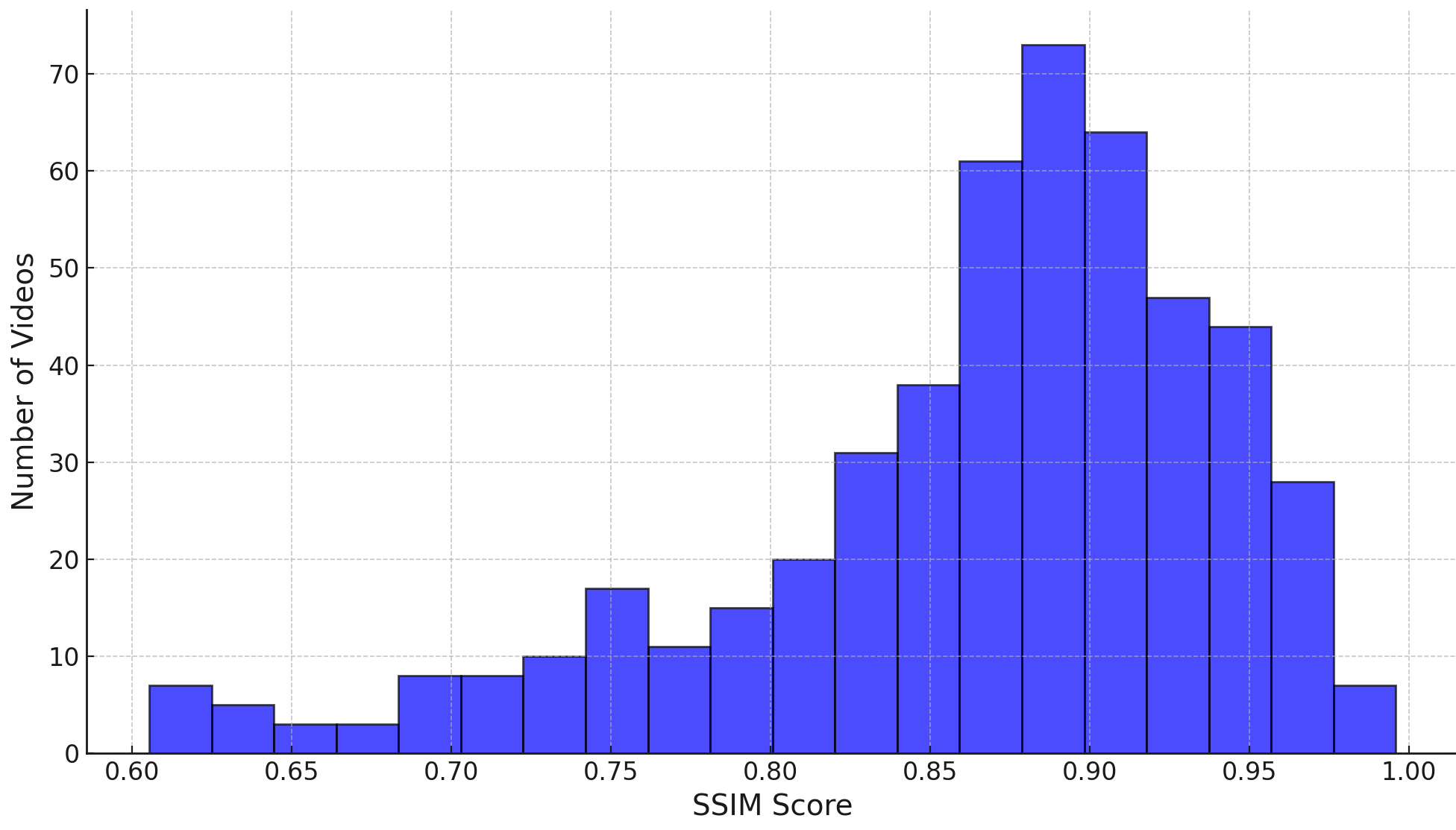} 
\caption{Distribution of SSIM scores for HAV-DF  dataset.}
\label{fig:SSIM}
\end{figure}
Figure~\ref{fig:SSIM} displays the distribution of structural similarity index measure (SSIM) scores for deepfake videos in the HAV-DF dataset. The x-axis represents the SSIM score, ranging from 0 to 1, while the y-axis shows the number of videos. Most of the videos have a very high SSIM score, clustered between 0.9 and 1.0, indicating a high degree of structural similarity between the deep-fake videos and their original counterparts. There is a significant spike in the number of videos with SSIM scores close to 1.0, suggesting that many deepfakes in this dataset are of very high quality and closely resemble the original videos. The majority of the scores in this dataset fall between 0.80 and 0.95, with the peak occurring around 0.85 to 0.90, suggesting that most videos exhibit a high level of structural similarity.
Lower scores (below 0.75) are relatively rare, indicating that only a small portion of the dataset contains videos with lower structural quality or distortions. This distribution suggests that the dataset primarily consists of high-quality or realistic videos, with only a minor subset showing reduced visual similarity.


\subsubsection{Comparison With Existing Methods: Video Datasets}
\begin{figure}[!ht]
    \centering
\includegraphics[width=0.75\linewidth]{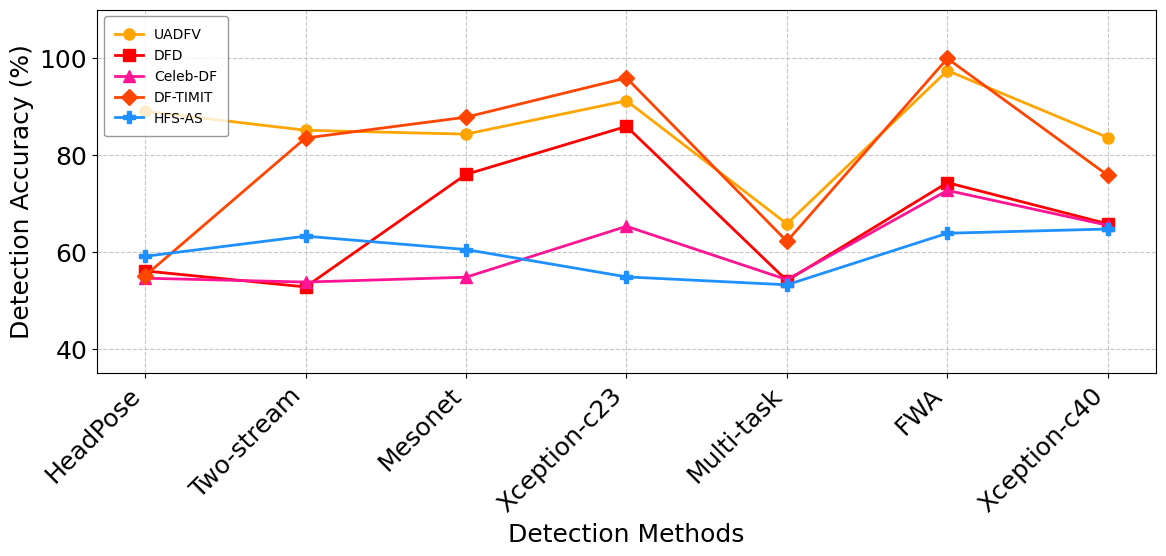} 

\caption{Comparative analysis of deepfake detection methods across multiple video datasets.}
\label{fig:comparison}
\end{figure}

Figure~\ref{fig:comparison} illustrates the performance of various deepfake detection methods across five different video datasets: UADFV~\citep{yang2019exposing}, DFDC~\citep{dufour2019deepfakes}, Celeb-DF~\citep{li2020celeb}, DF-TIMIT~\citep{korshunov2018deepfakes}, and HAV-DF (HindiFaceSwap-AudioSync). The x-axis represents different detection methods, including HeadPose~\citep{yang2019exposing}, Two-stream~\citep{zhou2017two}, Mesonet~\citep{afchar2018mesonet}, Xception-c23~\citep{chollet2017xception}, Multi-task~\citep{nguyen2019multi}, FWA~\citep{li2018exposing}, and Xception-c40~\citep{chollet2017xception}, while the y-axis shows the detection accuracy percentage.

The performance trends vary significantly across datasets and methods, revealing the complexity of deepfake detection. The DF-TIMIT dataset (pink line) generally shows the highest detection accuracies, 99.9\% with the FWA method. UADFV (orange line) also demonstrates consistently high performance across most methods. In contrast, the HAV-DF dataset (light blue line) consistently shows the lowest detection accuracies, rarely exceeding 65\%, suggesting that Hindi-language deepfakes in this dataset are particularly challenging to detect.
Interestingly, the graph reveals that different methods excel on different datasets. For instance, the Xception-c23 method performs exceptionally well on DF-TIMIT and UADFV but shows a significant drop in accuracy for other datasets, especially HAV-DF. The FWA method appears to be the most consistently high-performing across all datasets, with a notable peak for most datasets.
The Multi-task method shows the most variability, with a sharp dip in performance across all datasets, indicating it might be less reliable for general deepfake detection. The performance of the HAV-DF dataset remains consistently lower across all methods, highlighting the unique challenges posed by this Hindi-language dataset.

This comparative analysis underscores the importance of developing robust, cross-dataset deepfake detection methods. It also emphasizes the need for diverse, multilingual datasets in training and evaluating these methods to ensure their effectiveness across various types of deepfakes and cultural contexts. The consistently lower performance on the HAV-DF dataset suggests that current detection methods may need further refinement to adequately address the nuances of non-English language deepfakes (see Table~\ref{Table:comparison}).

\begin{table}[!ht]
\caption{Performance comparison of Hindi deep fake dataset (HAV-DF) with existing video datasets.}
\begin{tabular}{|l|c|c|c|c|c|}
\hline
\textbf{\begin{tabular}[c]{@{}l@{}}Datasets$\rightarrow$\\ Methods$\downarrow$ \end{tabular}} & \textbf{UADFV} & \textbf{DFD}  & \textbf{Celeb-DF} & \textbf{\begin{tabular}[c]{@{}c@{}}DF-TIMIT\\   	LQ\end{tabular}} & \textbf{HAV-DF} \\ \hline
HeadPose                                                            & 89.0           & 56.1          & \textbf{54.6}     & 55.1                                                              & 59.1           \\ \hline
Two-stream                                                   & 85.1           & \textbf{52.8} & 53.8              & 83.5                                                              & 63.3           \\ \hline
Mesonet                                                      & 84.3           & 76.0          & \textbf{54.8}     & 87.8                                                              & 60.5            \\ \hline
Xception-c23                                                 & 91.2           & 85.9          & 65.3              & 95.9                                                              & \textbf{54.8}   \\ \hline
Multi-task                                                   & 65.8           & 54.1          & 54.3              & 62.2                                                              & \textbf{53.2}   \\ \hline
FWA                                                         & 97.4           & 74.3          & 72.7              & 99.9                                                              & \textbf{63.8}   \\ \hline
Xception-c40                                                        & 83.6           & 65.8          & 65.5              & 75.8                                                              & \textbf{64.7}   \\ \hline
\end{tabular}
\label{Table:comparison}
\end{table}
\subsubsection{Comparison With Existing Methods: Audio-video Datasets}

\begin{figure}[!ht]
    \centering
\includegraphics[width=0.75\linewidth]{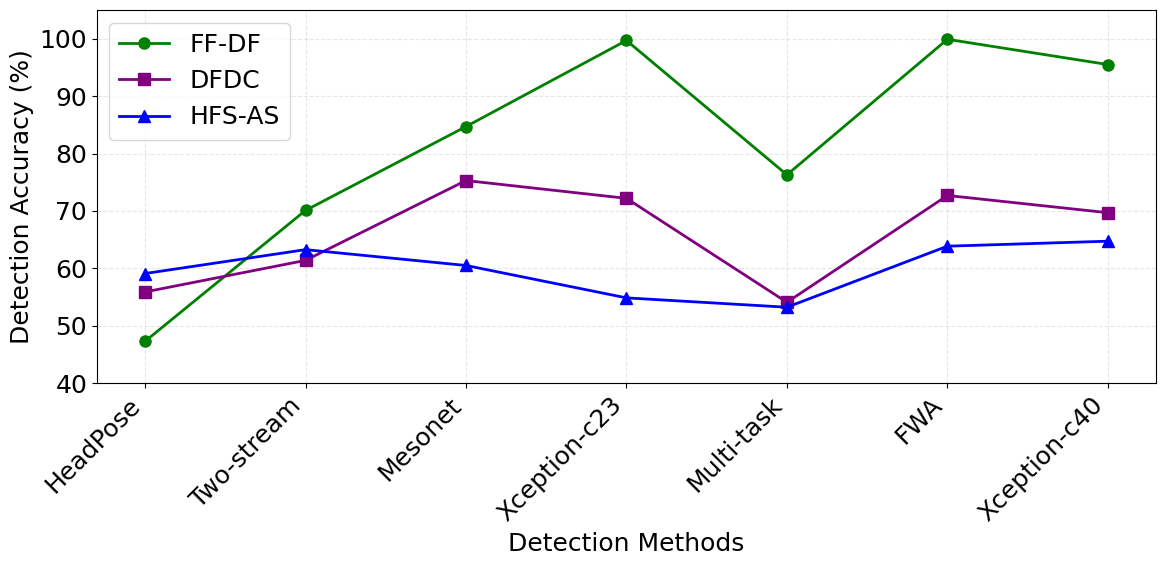} 

\caption{Comparative analysis of deepfake detection accuracy across different methods and audio-video datasets.}
\label{fig:Detection}
\end{figure}

Figure~\ref{fig:Detection} presents a comparative analysis of the accuracy of deepfake detection using different methods and English audio-video datasets. The graph plots detection accuracy percentages for three datasets: FF-DF (FaceForensics++), DFDC (DeepFake Detection Challenge) and HAV-DF. Seven detection methods are evaluated as Headpose, Two-stream, Mesonet, Xception-c23, Multi-task, FWA, and Xception-c40. The FF-DF dataset consistently shows the highest detection accuracy, peaking at 99.7\% with the Xception-c23 method. The DFDC data set generally performs second best, with accuracies fluctuating between 54.1\% and 75.3\%. In particular, the newly introduced HAV-DF dataset shows lower detection precisions in almost all methods except HeadPose, with its highest accuracy at 64.7\% using the Xception-c40 method. This trend suggests that the HAV-DF dataset presents deeper challenges to detect, possibly due to its focus on Hindi-language audio and diverse manipulation techniques. The graph effectively illustrates the varying difficulty levels in deepfake detection across different datasets and the relative performance of various detection methods, highlighting the complexities in identifying manipulated audio-visual content (see Table~\ref{table:audio-video}).
\begin{table}[!ht]
\caption{Performance comparison of Hindi deep fake dataset (HAV-DF) with existing audio-video datasets.}
\begin{tabular}{|l|c|c|c|}
\hline
\textbf{\begin{tabular}[c]{@{}l@{}}Datasets$\rightarrow$\\ Methods$\downarrow$ \end{tabular}} & \textbf{FF-DF} & \textbf{DFDC} & \textbf{HAV-DF} \\ \hline
HeadPose                                                            & \textbf{47.3}  & 55.9          & 59.1           \\ \hline
Two-stream                                                          & 70.1           & \textbf{61.4} & 63.3           \\ \hline
Mesonet                                                             & 84.7           & 75.3          & \textbf{60.5}   \\ \hline
Xception-c                                                          & 99.7           & 72.2          & \textbf{54.8}   \\ \hline
Multi-task                                                          & 76.3           & 54.1          & \textbf{53.2}   \\ \hline
FWA                                                                 & 99.9           & 72.7          & \textbf{63.8}   \\ \hline
Xception-c40                                                         & 95.5           & 69.7          & \textbf{64.7}   \\ \hline
\end{tabular}
\label{table:audio-video}
\end{table}
\section{Discussion and Conclusion}

This study introduces the HAV-DF dataset, a groundbreaking resource for deepfake research in the Hindi language. By addressing a critical gap in multilingual deepfake detection, HAV-DF provides a robust collection of 500 videos featuring face-swapping, lip-syncing, and voice-cloning techniques. Benchmarking against popular detection models reveals that HAV-DF poses significantly greater detection challenges than English-centric datasets, highlighting the complexity of Hindi-specific linguistic and cultural nuances. HAV-DF distinguishes itself by embracing linguistic diversity, integrating multimodal manipulations, and addressing detection challenges in low-resource settings, effectively filling critical gaps left by existing datasets. The potential of the HAV-DF dataset, as presented in this study, is outlined below.
\begin{enumerate}

    \item \textbf{Multimodel dataset:} HAV-DF dataset is multimodel as it uses audio as well as video combined deepfakes. As from the literature analysis, we could easily say that there is a gap for benchmark datasets for multimodel deepfakes. HAV-DF have shown less accuracy in multimodel deepfakes also as compared to other multimodel based datasets like FF-DF and DFDC. 

    \item \textbf{Language-specific focus:} HAV-DF is the first dataset specifically tailored for the Hindi language, addressing linguistic and cultural nuances often overlooked in existing datasets like DFD and Celeb-DF, which are primarily English-centric.

    \item \textbf{Comprehensive manipulations:} The dataset incorporates diverse deepfake techniques, including face-swapping, lip-syncing, and voice-cloning, enabling a more holistic evaluation of detection models compared to datasets that focus primarily on visual manipulations.
    
    \item \textbf{Low-resource language representation:} Unlike existing datasets like DFD and  Celeb-DF that predominantly cater to high-resource languages like English, HAV-DF highlights the challenges of deepfake detection in low-resource languages, filling a critical gap in multilingual research.

    \item \textbf{Detection difficulty:} Benchmarking results show that HAV-DF presents a higher level of difficulty for detection models like HeadPose, Two stream, Mesonet, Xception-c, Multi-task, FWA, and Xception-c40, making it a valuable resource for advancing detection technologies and improving model robustness.

    \item \textbf{Multimodal content:} By combining manipulated audio and video elements, HAV-DF supports research on multimodal deepfake detection, unlike many existing datasets that focus exclusively on either audio or video manipulations.

    
        
\end{enumerate}
\subsection{Limitations}
Although the HAV-DF dataset represents a significant milestone, several limitations exist:
\begin{itemize}
    \item \textbf{Dataset size:} The dataset is relatively small, comprising only 500 videos, limiting its utility for training large-scale detection models.
    
    \item \textbf{Diverse manipulation types:} As identified in the literature, a wide range of manipulation techniques exists for both video and audio. However, this study utilizes a limited set of manipulation techniques. In real-world scenarios, deepfake patterns are significantly more complex, involving intricate and varied manipulations in both modalities. Therefore, future work should focus on exploring additional manipulation techniques to better simulate real-life conditions and enhance the generation of multimodal deepfakes.
    
    \item \textbf{Multilingual deepfakes:} Due to globalization, every person uses more then one language for communication. The current research in audio deepfakes is concentrating more on monolingual audio. However, the future research should be focused more on exploring more than one languages during voice synthesis.

    \item \textbf{Face positioning constraint:} The dataset consists exclusively of videos featuring a single subject per frame, all captured in frontal views with subjects in a straight, upright position. This design reduces variability in pose and head angles, limiting the dataset's ability to represent realistic conditions where videos often include dynamic movements, multiple interacting faces, and non-frontal or occluded views. Such constraints may impact its applicability for real-world scenarios.

    \item \textbf{Linguistic and technical challenges:} Many deepfake generation tools and techniques are optimized for English or Western facial and vocal features, which poses significant challenges in creating realistic Hindi-language deepfakes. Additionally, the distinct sounds, intonations, and syllable structures of Hindi words differ markedly from English, making accurate and high-quality lip-syncing for Hindi particularly difficult. Hindi has numerous regional dialects (e.g., Awadhi, Bhojpuri, Bundeli), and it is challenging to incorporate all linguistic variations into a single dataset.
    
   

   
\end{itemize}


\section*{Data Availability Statement}
The datasets generated during and/or analysed during the current study along with python codes are available from the corresponding author on reasonable request.
%
%

 \section*{Conflict of interest}
 The author declares no conflict of interest.

\bibliographystyle{cas-model2-names}
\bibliography{cas-refs}

\begin{thebibliography}{45}
\expandafter\ifx\csname natexlab\endcsname\relax\def\natexlab#1{#1}\fi
\providecommand{\url}[1]{\texttt{#1}}
\providecommand{\href}[2]{#2}
\providecommand{\path}[1]{#1}
\providecommand{\DOIprefix}{doi:}
\providecommand{\ArXivprefix}{arXiv:}
\providecommand{\URLprefix}{URL: }
\providecommand{\Pubmedprefix}{pmid:}
\providecommand{\doi}[1]{\href{http://dx.doi.org/#1}{\path{#1}}}
\providecommand{\Pubmed}[1]{\href{pmid:#1}{\path{#1}}}
\providecommand{\bibinfo}[2]{#2}
\ifx\xfnm\relax \def\xfnm[#1]{\unskip,\space#1}\fi
\bibitem[{Afchar et~al.(2018)Afchar, Nozick, Yamagishi and Echizen}]{afchar2018mesonet}
\bibinfo{author}{Afchar, D.}, \bibinfo{author}{Nozick, V.}, \bibinfo{author}{Yamagishi, J.}, \bibinfo{author}{Echizen, I.}, \bibinfo{year}{2018}.
\newblock \bibinfo{title}{Mesonet: a compact facial video forgery detection network}, in: \bibinfo{booktitle}{2018 IEEE international workshop on information forensics and security (WIFS)}, \bibinfo{organization}{IEEE}. pp. \bibinfo{pages}{1--7}.
\bibitem[{Arik et~al.(2018)Arik, Chen, Peng, Ping and Zhou}]{arik2018neu}
\bibinfo{author}{Arik, S.}, \bibinfo{author}{Chen, J.}, \bibinfo{author}{Peng, K.}, \bibinfo{author}{Ping, W.}, \bibinfo{author}{Zhou, Y.}, \bibinfo{year}{2018}.
\newblock \bibinfo{title}{Neu-ralvoicecloning with a fewsamples. in advances in neural information processing systems}, in: \bibinfo{booktitle}{Annual Conference on Neural In formation Processing Systems 2018, Neur IPS 2018}, pp. \bibinfo{pages}{10040--10050}.
\bibitem[{Cheng et~al.(2022)Cheng, Cun, Zhang, Xia, Yin, Zhu, Wang, Wang and Wang}]{cheng2022videoretalking}
\bibinfo{author}{Cheng, K.}, \bibinfo{author}{Cun, X.}, \bibinfo{author}{Zhang, Y.}, \bibinfo{author}{Xia, M.}, \bibinfo{author}{Yin, F.}, \bibinfo{author}{Zhu, M.}, \bibinfo{author}{Wang, X.}, \bibinfo{author}{Wang, J.}, \bibinfo{author}{Wang, N.}, \bibinfo{year}{2022}.
\newblock \bibinfo{title}{Videoretalking: Audio-based lip synchronization for talking head video editing in the wild}, in: \bibinfo{booktitle}{SIGGRAPH Asia 2022 Conference Papers}, pp. \bibinfo{pages}{1--9}.
\bibitem[{Cheng(2024)}]{cheng2024refining}
\bibinfo{author}{Cheng, X.}, \bibinfo{year}{2024}.
\newblock \bibinfo{title}{Refining cyclegan with attention mechanisms and age-aware training for realistic deepfakes}.
\newblock \bibinfo{journal}{Heliyon} .
\bibitem[{Chesney and Citron(2019)}]{chesney2019deepfakes}
\bibinfo{author}{Chesney, R.}, \bibinfo{author}{Citron, D.}, \bibinfo{year}{2019}.
\newblock \bibinfo{title}{Deepfakes and the new disinformation war: The coming age of post-truth geopolitics}.
\newblock \bibinfo{journal}{Foreign Aff.} \bibinfo{volume}{98}, \bibinfo{pages}{147}.
\bibitem[{Chollet(2017)}]{chollet2017xception}
\bibinfo{author}{Chollet, F.}, \bibinfo{year}{2017}.
\newblock \bibinfo{title}{Xception: Deep learning with depthwise separable convolutions}, in: \bibinfo{booktitle}{Proceedings of the IEEE conference on computer vision and pattern recognition}, pp. \bibinfo{pages}{1251--1258}.
\bibitem[{Dang et~al.(2020)Dang, Liu, Stehouwer, Liu and Jain}]{dang2020detection}
\bibinfo{author}{Dang, H.}, \bibinfo{author}{Liu, F.}, \bibinfo{author}{Stehouwer, J.}, \bibinfo{author}{Liu, X.}, \bibinfo{author}{Jain, A.K.}, \bibinfo{year}{2020}.
\newblock \bibinfo{title}{On the detection of digital face manipulation}, in: \bibinfo{booktitle}{Proceedings of the IEEE/CVF Conference on Computer Vision and Pattern Recognition}, pp. \bibinfo{pages}{5781--5790}.
\bibitem[{Dolhansky et~al.(2020)Dolhansky, Bitton, Pflaum, Lu, Howes, Wang and Ferrer}]{dolhansky2020deepfake}
\bibinfo{author}{Dolhansky, B.}, \bibinfo{author}{Bitton, J.}, \bibinfo{author}{Pflaum, B.}, \bibinfo{author}{Lu, J.}, \bibinfo{author}{Howes, R.}, \bibinfo{author}{Wang, M.}, \bibinfo{author}{Ferrer, C.C.}, \bibinfo{year}{2020}.
\newblock \bibinfo{title}{The deepfake detection challenge (dfdc) dataset}.
\newblock \bibinfo{journal}{arXiv preprint arXiv:2006.07397} .
\bibitem[{Dufour et~al.(2019)Dufour, Gully, Karlsson, Vorbyov, Leung, Childs and Bregler}]{dufour2019deepfakes}
\bibinfo{author}{Dufour, N.}, \bibinfo{author}{Gully, A.}, \bibinfo{author}{Karlsson, P.}, \bibinfo{author}{Vorbyov, A.V.}, \bibinfo{author}{Leung, T.}, \bibinfo{author}{Childs, J.}, \bibinfo{author}{Bregler, C.}, \bibinfo{year}{2019}.
\newblock \bibinfo{title}{Deepfakes detection dataset by google \& jigsaw}, in: \bibinfo{booktitle}{arXiv preprint arXiv:1901.08971}.
\bibitem[{Felouat et~al.(2024)Felouat, Nguyen, Le, Yamagishi and Echizen}]{felouat2024ekyc}
\bibinfo{author}{Felouat, H.}, \bibinfo{author}{Nguyen, H.H.}, \bibinfo{author}{Le, T.N.}, \bibinfo{author}{Yamagishi, J.}, \bibinfo{author}{Echizen, I.}, \bibinfo{year}{2024}.
\newblock \bibinfo{title}{ekyc-df: A large-scale deepfake dataset for developing and evaluating ekyc systems}.
\newblock \bibinfo{journal}{IEEE Access} .
\bibitem[{He et~al.(2021)He, Gan, Chen, Zhou, Yin, Song, Sheng, Shao and Liu}]{he2021forgerynet}
\bibinfo{author}{He, Y.}, \bibinfo{author}{Gan, B.}, \bibinfo{author}{Chen, S.}, \bibinfo{author}{Zhou, Y.}, \bibinfo{author}{Yin, G.}, \bibinfo{author}{Song, L.}, \bibinfo{author}{Sheng, L.}, \bibinfo{author}{Shao, J.}, \bibinfo{author}{Liu, Z.}, \bibinfo{year}{2021}.
\newblock \bibinfo{title}{Forgerynet: A versatile benchmark for comprehensive forgery analysis}, in: \bibinfo{booktitle}{Proceedings of the IEEE/CVF conference on computer vision and pattern recognition}, pp. \bibinfo{pages}{4360--4369}.
\bibitem[{Hou et~al.(2024)Hou, Fu, Chen, Li, Zhang and Zhao}]{hou2024polyglotfake}
\bibinfo{author}{Hou, Y.}, \bibinfo{author}{Fu, H.}, \bibinfo{author}{Chen, C.}, \bibinfo{author}{Li, Z.}, \bibinfo{author}{Zhang, H.}, \bibinfo{author}{Zhao, J.}, \bibinfo{year}{2024}.
\newblock \bibinfo{title}{Polyglotfake: A novel multilingual and multimodal deepfake dataset}.
\newblock \bibinfo{journal}{arXiv preprint arXiv:2405.08838} .
\bibitem[{Jia et~al.(2018)Jia, Zhang, Weiss, Wang, Shen, Ren, Nguyen, Pang, Lopez~Moreno, Wu et~al.}]{jia2018transfer}
\bibinfo{author}{Jia, Y.}, \bibinfo{author}{Zhang, Y.}, \bibinfo{author}{Weiss, R.}, \bibinfo{author}{Wang, Q.}, \bibinfo{author}{Shen, J.}, \bibinfo{author}{Ren, F.}, \bibinfo{author}{Nguyen, P.}, \bibinfo{author}{Pang, R.}, \bibinfo{author}{Lopez~Moreno, I.}, \bibinfo{author}{Wu, Y.}, et~al., \bibinfo{year}{2018}.
\newblock \bibinfo{title}{Transfer learning from speaker verification to multispeaker text-to-speech synthesis}.
\newblock \bibinfo{journal}{Advances in neural information processing systems} \bibinfo{volume}{31}.
\bibitem[{Jiang et~al.(2020)Jiang, Li, Wu, Qian and Loy}]{jiang2020deeperforensics}
\bibinfo{author}{Jiang, L.}, \bibinfo{author}{Li, R.}, \bibinfo{author}{Wu, W.}, \bibinfo{author}{Qian, C.}, \bibinfo{author}{Loy, C.C.}, \bibinfo{year}{2020}.
\newblock \bibinfo{title}{Deeperforensics-1.0: A large-scale dataset for real-world face forgery detection}.
\newblock \bibinfo{journal}{arXiv preprint arXiv:2001.03024} .
\bibitem[{Kambali et~al.(2023)Kambali, Ansari, Srivastav, Aryan and Nanda}]{kambali2023real}
\bibinfo{author}{Kambali, S.P.}, \bibinfo{author}{Ansari, M.A.}, \bibinfo{author}{Srivastav, P.U.}, \bibinfo{author}{Aryan, M.D.}, \bibinfo{author}{Nanda, R.}, \bibinfo{year}{2023}.
\newblock \bibinfo{title}{Real time voice cloning system}.
\newblock \bibinfo{journal}{International Research Journal of Innovations in Engineering and Technology} \bibinfo{volume}{7}, \bibinfo{pages}{294}.
\bibitem[{Khalid et~al.(2021)Khalid, Tariq, Kim and Woo}]{khalid2021fakeavceleb}
\bibinfo{author}{Khalid, H.}, \bibinfo{author}{Tariq, S.}, \bibinfo{author}{Kim, M.}, \bibinfo{author}{Woo, S.S.}, \bibinfo{year}{2021}.
\newblock \bibinfo{title}{Fakeavceleb: A novel audio-video multimodal deepfake dataset}.
\newblock \bibinfo{journal}{arXiv preprint arXiv:2108.05080} .
\bibitem[{Korshunov and Marcel(2018)}]{korshunov2018deepfakes}
\bibinfo{author}{Korshunov, P.}, \bibinfo{author}{Marcel, S.}, \bibinfo{year}{2018}.
\newblock \bibinfo{title}{Deepfakes: a new threat to face recognition? assessment and detection}.
\newblock \bibinfo{journal}{arXiv preprint arXiv:1812.08685} .
\bibitem[{Kwon et~al.(2021)Kwon, You, Nam, Park and Chae}]{kwon2021kodf}
\bibinfo{author}{Kwon, P.}, \bibinfo{author}{You, J.}, \bibinfo{author}{Nam, G.}, \bibinfo{author}{Park, S.}, \bibinfo{author}{Chae, G.}, \bibinfo{year}{2021}.
\newblock \bibinfo{title}{Kodf: A large-scale korean deepfake detection dataset}, in: \bibinfo{booktitle}{Proceedings of the IEEE/CVF international conference on computer vision}, pp. \bibinfo{pages}{10744--10753}.
\bibitem[{Li(2018)}]{li2018exposing}
\bibinfo{author}{Li, Y.}, \bibinfo{year}{2018}.
\newblock \bibinfo{title}{Exposing deepfake videos by detecting face warping artif acts}.
\newblock \bibinfo{journal}{arXiv preprint arXiv:1811.00656} .
\bibitem[{Li et~al.(2020)Li, Yang, Sun, Qi and Lyu}]{li2020celeb}
\bibinfo{author}{Li, Y.}, \bibinfo{author}{Yang, X.}, \bibinfo{author}{Sun, P.}, \bibinfo{author}{Qi, H.}, \bibinfo{author}{Lyu, S.}, \bibinfo{year}{2020}.
\newblock \bibinfo{title}{Celeb-df: A large-scale challenging dataset for deepfake forensics}, in: \bibinfo{booktitle}{Proceedings of the IEEE/CVF conference on computer vision and pattern recognition}, pp. \bibinfo{pages}{3207--3216}.
\bibitem[{Liu et~al.(2023)Liu, Perov, Gao, Chervoniy, Zhou and Zhang}]{liu2023deepfacelab}
\bibinfo{author}{Liu, K.}, \bibinfo{author}{Perov, I.}, \bibinfo{author}{Gao, D.}, \bibinfo{author}{Chervoniy, N.}, \bibinfo{author}{Zhou, W.}, \bibinfo{author}{Zhang, W.}, \bibinfo{year}{2023}.
\newblock \bibinfo{title}{Deepfacelab: Integrated, flexible and extensible face-swapping framework}.
\newblock \bibinfo{journal}{Pattern Recognition} \bibinfo{volume}{141}, \bibinfo{pages}{109628}.
\bibitem[{Liz-Lopez et~al.(2024)Liz-Lopez, Keita, Taleb-Ahmed, Hadid, Huertas-Tato and Camacho}]{liz2024generation}
\bibinfo{author}{Liz-Lopez, H.}, \bibinfo{author}{Keita, M.}, \bibinfo{author}{Taleb-Ahmed, A.}, \bibinfo{author}{Hadid, A.}, \bibinfo{author}{Huertas-Tato, J.}, \bibinfo{author}{Camacho, D.}, \bibinfo{year}{2024}.
\newblock \bibinfo{title}{Generation and detection of manipulated multimodal audiovisual content: Advances, trends and open challenges}.
\newblock \bibinfo{journal}{Information Fusion} \bibinfo{volume}{103}, \bibinfo{pages}{102103}.
\bibitem[{Masood et~al.(2023)Masood, Nawaz, Malik, Javed, Irtaza and Malik}]{masood2023deepfakes}
\bibinfo{author}{Masood, M.}, \bibinfo{author}{Nawaz, M.}, \bibinfo{author}{Malik, K.M.}, \bibinfo{author}{Javed, A.}, \bibinfo{author}{Irtaza, A.}, \bibinfo{author}{Malik, H.}, \bibinfo{year}{2023}.
\newblock \bibinfo{title}{Deepfakes generation and detection: State-of-the-art, open challenges, countermeasures, and way forward}.
\newblock \bibinfo{journal}{Applied intelligence} \bibinfo{volume}{53}, \bibinfo{pages}{3974--4026}.
\bibitem[{Melnik et~al.(2024)Melnik, Miasayedzenkau, Makaravets, Pirshtuk, Akbulut, Holzmann, Renusch, Reichert and Ritter}]{melnik2024face}
\bibinfo{author}{Melnik, A.}, \bibinfo{author}{Miasayedzenkau, M.}, \bibinfo{author}{Makaravets, D.}, \bibinfo{author}{Pirshtuk, D.}, \bibinfo{author}{Akbulut, E.}, \bibinfo{author}{Holzmann, D.}, \bibinfo{author}{Renusch, T.}, \bibinfo{author}{Reichert, G.}, \bibinfo{author}{Ritter, H.}, \bibinfo{year}{2024}.
\newblock \bibinfo{title}{Face generation and editing with stylegan: A survey}.
\newblock \bibinfo{journal}{IEEE Transactions on Pattern Analysis and Machine Intelligence} .
\bibitem[{Mirsky and Lee(2021)}]{mirsky2021creation}
\bibinfo{author}{Mirsky, Y.}, \bibinfo{author}{Lee, W.}, \bibinfo{year}{2021}.
\newblock \bibinfo{title}{The creation and detection of deepfakes: A survey}.
\newblock \bibinfo{journal}{ACM computing surveys (CSUR)} \bibinfo{volume}{54}, \bibinfo{pages}{1--41}.
\bibitem[{Mubarak et~al.(2023)Mubarak, Alsboui, Alshaikh, Inuwa-Dute, Khan and Parkinson}]{mubarak2023survey}
\bibinfo{author}{Mubarak, R.}, \bibinfo{author}{Alsboui, T.}, \bibinfo{author}{Alshaikh, O.}, \bibinfo{author}{Inuwa-Dute, I.}, \bibinfo{author}{Khan, S.}, \bibinfo{author}{Parkinson, S.}, \bibinfo{year}{2023}.
\newblock \bibinfo{title}{A survey on the detection and impacts of deepfakes in visual, audio, and textual formats}.
\newblock \bibinfo{journal}{IEEE Access} .
\bibitem[{Munir et~al.(2024)Munir, Sajjad, Raza, Abbas, Azeemi, Qazi and Raza}]{munir2024deepfake}
\bibinfo{author}{Munir, S.}, \bibinfo{author}{Sajjad, W.}, \bibinfo{author}{Raza, M.}, \bibinfo{author}{Abbas, E.}, \bibinfo{author}{Azeemi, A.H.}, \bibinfo{author}{Qazi, I.A.}, \bibinfo{author}{Raza, A.A.}, \bibinfo{year}{2024}.
\newblock \bibinfo{title}{Deepfake defense: Constructing and evaluating a specialized urdu deepfake audio dataset}, in: \bibinfo{booktitle}{Findings of the Association for Computational Linguistics ACL 2024}, pp. \bibinfo{pages}{14470--14480}.
\bibitem[{Narayan et~al.(2023)Narayan, Agarwal, Thakral, Mittal, Vatsa and Singh}]{narayan2023df}
\bibinfo{author}{Narayan, K.}, \bibinfo{author}{Agarwal, H.}, \bibinfo{author}{Thakral, K.}, \bibinfo{author}{Mittal, S.}, \bibinfo{author}{Vatsa, M.}, \bibinfo{author}{Singh, R.}, \bibinfo{year}{2023}.
\newblock \bibinfo{title}{Df-platter: Multi-face heterogeneous deepfake dataset}, in: \bibinfo{booktitle}{Proceedings of the IEEE/CVF Conference on Computer Vision and Pattern Recognition}, pp. \bibinfo{pages}{9739--9748}.
\bibitem[{Nautsch et~al.(2021)Nautsch, Wang, Evans, Kinnunen, Vestman, Todisco, Delgado, Sahidullah, Yamagishi and Lee}]{nautsch2021asvspoof}
\bibinfo{author}{Nautsch, A.}, \bibinfo{author}{Wang, X.}, \bibinfo{author}{Evans, N.}, \bibinfo{author}{Kinnunen, T.H.}, \bibinfo{author}{Vestman, V.}, \bibinfo{author}{Todisco, M.}, \bibinfo{author}{Delgado, H.}, \bibinfo{author}{Sahidullah, M.}, \bibinfo{author}{Yamagishi, J.}, \bibinfo{author}{Lee, K.A.}, \bibinfo{year}{2021}.
\newblock \bibinfo{title}{Asvspoof 2019: spoofing countermeasures for the detection of synthesized, converted and replayed speech}.
\newblock \bibinfo{journal}{IEEE Transactions on Biometrics, Behavior, and Identity Science} \bibinfo{volume}{3}, \bibinfo{pages}{252--265}.
\bibitem[{Nguyen et~al.(2019)Nguyen, Fang, Yamagishi and Echizen}]{nguyen2019multi}
\bibinfo{author}{Nguyen, H.H.}, \bibinfo{author}{Fang, F.}, \bibinfo{author}{Yamagishi, J.}, \bibinfo{author}{Echizen, I.}, \bibinfo{year}{2019}.
\newblock \bibinfo{title}{Multi-task learning for detecting and segmenting manipulated facial images and videos}, in: \bibinfo{booktitle}{2019 IEEE 10th International Conference on Biometrics Theory, Applications and Systems (BTAS)}, \bibinfo{organization}{IEEE}. pp. \bibinfo{pages}{1--8}.
\bibitem[{Nirkin et~al.(2019)Nirkin, Keller and Hassner}]{nirkin2019fsgan}
\bibinfo{author}{Nirkin, Y.}, \bibinfo{author}{Keller, Y.}, \bibinfo{author}{Hassner, T.}, \bibinfo{year}{2019}.
\newblock \bibinfo{title}{Fsgan: Subject agnostic face swapping and reenactment}, in: \bibinfo{booktitle}{Proceedings of the IEEE/CVF international conference on computer vision}, pp. \bibinfo{pages}{7184--7193}.
\bibitem[{Patel et~al.(2023)Patel, Tanwar, Gupta, Bhattacharya, Davidson, Nyameko, Aluvala and Vimal}]{patel2023deepfake}
\bibinfo{author}{Patel, Y.}, \bibinfo{author}{Tanwar, S.}, \bibinfo{author}{Gupta, R.}, \bibinfo{author}{Bhattacharya, P.}, \bibinfo{author}{Davidson, I.E.}, \bibinfo{author}{Nyameko, R.}, \bibinfo{author}{Aluvala, S.}, \bibinfo{author}{Vimal, V.}, \bibinfo{year}{2023}.
\newblock \bibinfo{title}{Deepfake generation and detection: Case study and challenges}.
\newblock \bibinfo{journal}{IEEE Access} .
\bibitem[{Perov et~al.(2020)Perov, Gao, Chervoniy, Liu, Marangonda, Um{\'e}, Dpfks, Facenheim, RP, Jiang et~al.}]{perov2020deepfacelab}
\bibinfo{author}{Perov, I.}, \bibinfo{author}{Gao, D.}, \bibinfo{author}{Chervoniy, N.}, \bibinfo{author}{Liu, K.}, \bibinfo{author}{Marangonda, S.}, \bibinfo{author}{Um{\'e}, C.}, \bibinfo{author}{Dpfks, M.}, \bibinfo{author}{Facenheim, C.S.}, \bibinfo{author}{RP, L.}, \bibinfo{author}{Jiang, J.}, et~al., \bibinfo{year}{2020}.
\newblock \bibinfo{title}{Deepfacelab: Integrated, flexible and extensible face-swapping framework}.
\newblock \bibinfo{journal}{arXiv preprint arXiv:2005.05535} .
\bibitem[{Rabhi et~al.(2024)Rabhi, Bakiras and Di~Pietro}]{rabhi2024audio}
\bibinfo{author}{Rabhi, M.}, \bibinfo{author}{Bakiras, S.}, \bibinfo{author}{Di~Pietro, R.}, \bibinfo{year}{2024}.
\newblock \bibinfo{title}{Audio-deepfake detection: Adversarial attacks and countermeasures}.
\newblock \bibinfo{journal}{Expert Systems with Applications} \bibinfo{volume}{250}, \bibinfo{pages}{123941}.
\bibitem[{Rossler et~al.(2019)Rossler, Cozzolino, Verdoliva, Riess, Thies and Nie{\ss}ner}]{rossler2019faceforensics++}
\bibinfo{author}{Rossler, A.}, \bibinfo{author}{Cozzolino, D.}, \bibinfo{author}{Verdoliva, L.}, \bibinfo{author}{Riess, C.}, \bibinfo{author}{Thies, J.}, \bibinfo{author}{Nie{\ss}ner, M.}, \bibinfo{year}{2019}.
\newblock \bibinfo{title}{Faceforensics++: Learning to detect manipulated facial images}, in: \bibinfo{booktitle}{Proceedings of the IEEE/CVF international conference on computer vision}, pp. \bibinfo{pages}{1--11}.
\bibitem[{Seong et~al.(2021)Seong, Lee and Lee}]{seong2021multilingual}
\bibinfo{author}{Seong, J.}, \bibinfo{author}{Lee, W.}, \bibinfo{author}{Lee, S.}, \bibinfo{year}{2021}.
\newblock \bibinfo{title}{Multilingual speech synthesis for voice cloning}, in: \bibinfo{booktitle}{2021 IEEE International Conference on Big Data and Smart Computing (BigComp)}, \bibinfo{organization}{IEEE}. pp. \bibinfo{pages}{313--316}.
\bibitem[{Shelar et~al.(2022)Shelar, Ghatole, Pachpande, Bhandari and Shinde}]{shelar2022deepfakes}
\bibinfo{author}{Shelar, J.}, \bibinfo{author}{Ghatole, D.}, \bibinfo{author}{Pachpande, M.}, \bibinfo{author}{Bhandari, D.}, \bibinfo{author}{Shinde, S.}, \bibinfo{year}{2022}.
\newblock \bibinfo{title}{Deepfakes for video conferencing using general adversarial networks (gans) and multilingual voice cloning}, in: \bibinfo{booktitle}{Computational Intelligence in Data Mining: Proceedings of ICCIDM 2021}. \bibinfo{publisher}{Springer}, pp. \bibinfo{pages}{137--148}.
\bibitem[{Shen et~al.(2018)Shen, Pang, Weiss, Schuster, Jaitly, Yang, Chen, Zhang, Wang, Skerrv-Ryan et~al.}]{shen2018natural}
\bibinfo{author}{Shen, J.}, \bibinfo{author}{Pang, R.}, \bibinfo{author}{Weiss, R.J.}, \bibinfo{author}{Schuster, M.}, \bibinfo{author}{Jaitly, N.}, \bibinfo{author}{Yang, Z.}, \bibinfo{author}{Chen, Z.}, \bibinfo{author}{Zhang, Y.}, \bibinfo{author}{Wang, Y.}, \bibinfo{author}{Skerrv-Ryan, R.}, et~al., \bibinfo{year}{2018}.
\newblock \bibinfo{title}{Natural tts synthesis by conditioning wavenet on mel spectrogram predictions}, in: \bibinfo{booktitle}{2018 IEEE international conference on acoustics, speech and signal processing (ICASSP)}, \bibinfo{organization}{IEEE}. pp. \bibinfo{pages}{4779--4783}.
\bibitem[{Tolosana et~al.(2020)Tolosana, Vera-Rodriguez, Fierrez, Morales and Ortega-Garcia}]{tolosana2020deepfakes}
\bibinfo{author}{Tolosana, R.}, \bibinfo{author}{Vera-Rodriguez, R.}, \bibinfo{author}{Fierrez, J.}, \bibinfo{author}{Morales, A.}, \bibinfo{author}{Ortega-Garcia, J.}, \bibinfo{year}{2020}.
\newblock \bibinfo{title}{Deepfakes and beyond: A survey of face manipulation and fake detection}.
\newblock \bibinfo{journal}{Information Fusion} \bibinfo{volume}{64}, \bibinfo{pages}{131--148}.
\bibitem[{Wang et~al.(2017)Wang, Skerry-Ryan, Stanton, Wu, Weiss, Jaitly, Yang, Xiao, Chen, Bengio et~al.}]{wang2017tacotron}
\bibinfo{author}{Wang, Y.}, \bibinfo{author}{Skerry-Ryan, R.}, \bibinfo{author}{Stanton, D.}, \bibinfo{author}{Wu, Y.}, \bibinfo{author}{Weiss, R.J.}, \bibinfo{author}{Jaitly, N.}, \bibinfo{author}{Yang, Z.}, \bibinfo{author}{Xiao, Y.}, \bibinfo{author}{Chen, Z.}, \bibinfo{author}{Bengio, S.}, et~al., \bibinfo{year}{2017}.
\newblock \bibinfo{title}{Tacotron: Towards end-to-end speech synthesis}.
\newblock \bibinfo{journal}{arXiv preprint arXiv:1703.10135} .
\bibitem[{Wu et~al.(2017)Wu, Yamagishi, Kinnunen, Hanil{\c{c}}i, Sahidullah, Sizov, Evans, Todisco and Delgado}]{wu2017asvspoof}
\bibinfo{author}{Wu, Z.}, \bibinfo{author}{Yamagishi, J.}, \bibinfo{author}{Kinnunen, T.}, \bibinfo{author}{Hanil{\c{c}}i, C.}, \bibinfo{author}{Sahidullah, M.}, \bibinfo{author}{Sizov, A.}, \bibinfo{author}{Evans, N.}, \bibinfo{author}{Todisco, M.}, \bibinfo{author}{Delgado, H.}, \bibinfo{year}{2017}.
\newblock \bibinfo{title}{Asvspoof: the automatic speaker verification spoofing and countermeasures challenge}.
\newblock \bibinfo{journal}{IEEE Journal of Selected Topics in Signal Processing} \bibinfo{volume}{11}, \bibinfo{pages}{588--604}.
\bibitem[{Xie et~al.(2020)Xie, Chatterjee, Liu, Roy and Kossi}]{xie2020deepfake}
\bibinfo{author}{Xie, D.}, \bibinfo{author}{Chatterjee, P.}, \bibinfo{author}{Liu, Z.}, \bibinfo{author}{Roy, K.}, \bibinfo{author}{Kossi, E.}, \bibinfo{year}{2020}.
\newblock \bibinfo{title}{Deepfake detection on publicly available datasets using modified alexnet}, in: \bibinfo{booktitle}{2020 IEEE Symposium Series on Computational Intelligence (SSCI)}, \bibinfo{organization}{IEEE}. pp. \bibinfo{pages}{1866--1871}.
\bibitem[{Yang et~al.(2019)Yang, Li and Lyu}]{yang2019exposing}
\bibinfo{author}{Yang, X.}, \bibinfo{author}{Li, Y.}, \bibinfo{author}{Lyu, S.}, \bibinfo{year}{2019}.
\newblock \bibinfo{title}{Exposing deep fakes using inconsistent head poses}, in: \bibinfo{booktitle}{ICASSP 2019-2019 IEEE International Conference on Acoustics, Speech and Signal Processing (ICASSP)}, \bibinfo{organization}{IEEE}. pp. \bibinfo{pages}{8261--8265}.
\bibitem[{Zhou et~al.(2017)Zhou, Han, Morariu and Davis}]{zhou2017two}
\bibinfo{author}{Zhou, P.}, \bibinfo{author}{Han, X.}, \bibinfo{author}{Morariu, V.I.}, \bibinfo{author}{Davis, L.S.}, \bibinfo{year}{2017}.
\newblock \bibinfo{title}{Two-stream neural networks for tampered face detection}, in: \bibinfo{booktitle}{2017 IEEE Conference on Computer Vision and Pattern Recognition Workshops (CVPRW)}, \bibinfo{organization}{IEEE}. pp. \bibinfo{pages}{1831--1839}.
\bibitem[{Zi et~al.(2020)Zi, Chang, Chen, Ma and Jiang}]{zi2020wilddeepfake}
\bibinfo{author}{Zi, B.}, \bibinfo{author}{Chang, M.}, \bibinfo{author}{Chen, J.}, \bibinfo{author}{Ma, X.}, \bibinfo{author}{Jiang, Y.G.}, \bibinfo{year}{2020}.
\newblock \bibinfo{title}{Wilddeepfake: A challenging real-world dataset for deepfake detection}, in: \bibinfo{booktitle}{Proceedings of the 28th ACM international conference on multimedia}, pp. \bibinfo{pages}{2382--2390}.

\end{thebibliography}

\end{document}